\RequirePackage{ifpdf}
\documentclass[hyper,letterpaper]{JHEP3}
\pdfoutput=1
\usepackage{amsmath,amssymb,amsfonts,bm,empheq}
\usepackage{cite}
\usepackage{amsmath,amsthm}
\usepackage{nicefrac}
\usepackage[enableskew,vcentermath]{youngtab}

\usepackage{float}

\def\printname#1{
        \if\draft y
                \smash{\makebox[0pt]{\hspace{-0.5in}
                        \raisebox{8pt}{\tt\tiny #1}}}
        \fi
}

\newlength{\standardunitlength}
\catcode`\@=11
\long\def\@makecaption#1#2{%
     \vskip 10pt

\setbox\@tempboxa\hbox{
       \small\sf{\bfcaptionfont #1. }\ignorespaces #2}%
     \ifdim \wd\@tempboxa >\captionwidth {%
         \rightskip=\@captionmargin\leftskip=\@captionmargin
         \unhbox\@tempboxa\par}%
       \else
         \hbox to\hsize{\hfil\box\@tempboxa\hfil}%
     \fi}
\font\bfcaptionfont=cmssbx10 scaled \magstephalf
\newdimen\@captionmargin\@captionmargin=2\parindent
\newdimen\captionwidth\captionwidth=\hsize
\catcode`\@=12

\newcommand{\bea}{\begin{eqnarray}}
\newcommand{\eea}{\end{eqnarray}}
\newcommand{\be}{\begin{equation}}
\newcommand{\ee}{\end{equation}}

\def\Tr{{\rm Tr \,}}

\def\k{\kappa}

\def\r{\rho}

\renewcommand{\bar}{\overline}
\renewcommand{\hat}{\widehat}

\newcommand\catalannumber[3]{
  \fill[white]  (#1) rectangle +(#2,#2);
  \fill[fill=gray!25]
  (#1)
  \foreach \dir in {#3}{
    \ifnum\dir=0
    -- ++(0,1)
    \else
    -- ++(1,0)
    \fi
  } |- (#1);
  \draw[help lines] (#1) grid +(#2,#2);
  \draw[dashed] (#1) -- +(#2,#2);
  \coordinate (prev) at (#1);
  \foreach \dir in {#3}{
    \ifnum\dir=0
    \coordinate (dep) at (0,1);
    \else
    \coordinate (dep) at (1,0);
    \fi
    \draw[line width=2pt,-stealth] (prev) -- ++(dep) coordinate (prev);
  };
}

\newcommand{\alphabar}{{\bar{\alpha}}}

\title{BPS counting for knots and combinatorics on words}

\author{Piotr Kucharski$^{1}$ and Piotr Su{\l}kowski$^{1,2}$
\\
$^1$ Faculty of Physics, University of Warsaw, ul. Pasteura 5, 02-093 Warsaw, Poland \\
$^2$ Walter Burke Institute for Theoretical Physics, California Institute of Technology, Pasadena, CA 91125, USA 
}

\abstract{We discuss relations between quantum BPS invariants defined in terms of a product decomposition of certain series, and difference equations (quantum A-polynomials) that annihilate such series. We construct combinatorial models whose structure is encoded in the form of such difference equations, and whose generating functions (Hilbert-Poincar{\'e} series) are solutions to those equations and reproduce generating series that encode BPS invariants. Furthermore, BPS invariants in question are expressed in terms of Lyndon words in an appropriate language, thereby relating counting of BPS states to the branch of mathematics referred to as combinatorics on words. We illustrate these results in the framework of colored extremal knot polynomials: among others we determine dual quantum extremal A-polynomials for various knots, present associated combinatorial models, find corresponding BPS invariants (extremal Labastida-Mari{\~n}o-Ooguri-Vafa invariants) and discuss their integrality.
\\
\\
\\
\\
\\
\\
\\
\\
\\
\\
\\
{\tt CALT-2016-022}}

\begin{document}



\newpage

\section{Introduction}
\label{sec.intro}

Counting of BPS states provides an important information about supersymmetric theories and has led to important advances in high energy physics and mathematical physics. In this paper we present a universal construction of combinatorial models related to the counting of a certain class of BPS states. While BPS counting is related to numerous mathematical fields, our discussion on one hand focuses on the issues of quantum curves and A-polynomials, and on the other hand it reveals intimate links of BPS counting with a relatively new area of discrete mathematics, referred to as combinatorics on words \cite{Lothaire1983,Lothaire2002,Berstel2007996}. 

There are certain classes of BPS invariants, which are defined in terms of a product decomposition of some generating series. One example of such invariants are Gopakumar-Vafa invariants considered in the context of closed topological string theory \cite{Gopakumar:1998ii,Gopakumar:1998jq}. Analogous invariants for open topological strings were discussed in \cite{AV-discs,AKV-framing}, and in particular they were related to knots in  \cite{OoguriV,Labastida:2000zp,Labastida:2000yw}. Integrality of BPS invariants related to topological strings was subsequently discussed among others in \cite{Kontsevich:2006an,Vologodsky:2007ef,Schwarz:2008ti,Schwarz:2013zua}. In mathematics invariants defined in terms of a product decomposition arise also in Donaldson-Thomas theory. A general theory of Donaldson-Thomas invariants was formulated in \cite{Kontsevich:2008fj}, and its physical interpretations have been discussed among others in \cite{Gaiotto:2008cd,Dimofte:2009bv}. Donaldson-Thomas invariants defined in terms of product decompositions of certain series have been analyzed in particular in \cite{COM:8276935,Rei12}. There are two classes of all above mentioned BPS invariants, referred to as classical and quantum. The definition of the latter ones, also called refined or motivic, involves an additional parameter $q$, such that the classical invariants are recovered in the $q\to 1$ limit. 

While our results are of more general interest, the analysis in this paper is conducted primarily in the context of Labastida-Mari{\~n}o-Ooguri-Vafa (LMOV) invariants associated to knots \cite{OoguriV,Labastida:2000zp,Labastida:2000yw}. From physics perspective LMOV invariants count the number of M2-branes attached to M5-branes in the conifold geometry. The three-dimensional part of M5-branes spans a lagrangian submanifold in the conifold, whose geometry is determined by a type of a knot. LMOV invariants can be regarded as a reformulation of colored HOMFLY polynomials $P_R(a,q)$, which are labeled by arbitrary representations (Young diagrams) $R$ and depend on two parameters $a$ and $q$. In order to determine LMOV invariants one needs to combine colored HOMFLY polynomials into a generating series and consider its product decomposition, with the argument $q$ of HOMFLY polynomials identified as the quantum parameter.

HOMFLY polynomials $P_r(a,q)\equiv P_{S^r}(a,q)$ labeled by symmetric representations $R=S^r$ form an interesting class \cite{Gukov:2011ry,Wedrich:2014zua,Garoufalidis:2015ewa}. On one hand, it is known that such polynomials satisfy recursion relations that can be represented in terms of generalized quantum A-polynomials \cite{AVqdef,superA,Nawata,FGSS,Garoufalidis:2016zhf,Gukov:2015gmm}, closely related to augmentation polynomials \cite{Ng}. On the other hand, they form a closed subsystem, within which LMOV invariants can be consistently defined \cite{Garoufalidis:2015ewa}. Therefore the structure of this class of LMOV invariants should be encoded in quantum A-polynomials, and one aim of this work is to reveal such a connection. Moreover, in the classical limit $q \to 1$ quantum A-polynomials reduce to classical algebraic curves, and it was shown in \cite{Garoufalidis:2015ewa} that such algebraic curves indeed encode classical LMOV invariants. Our present work can be therefore regarded as a generalization of \cite{Garoufalidis:2015ewa} to the quantum case. As in \cite{Garoufalidis:2015ewa}, in this work we also introduce one additional simplification and consider extremal HOMFLY polynomials, namely coefficients of the highest or lowest powers of $a$ in a given colored HOMFLY polynomial, which we denote respectively as $P^{\pm}_r(q)$, or simply $P_r(q)$. One advantage of the analysis of extremal polynomials is a chance of obtaining explicit, exact results that represent main features of a problem, without delving into technicalities. We denote the corresponding extremal LMOV invariants as $N^{\pm}_{r,j}$ or simply $N_{r,j}$.

Note that (extremal) quantum A-polynomials are examples of quantum curves, which are objects that have been actively studied in last years \cite{abmodel,Norbury-quantum,Schwarz:2014hfa,Dumitrescu:2015mpa,Manabe:2015kbj,Bouchard:2016obz}. One interesting problem in this field is how to determine whether a given classical algebraic curve is quantizable, and how to formulate a general quantization procedure, which lifts such an algebraic curve into a quantum curve. We believe that the relation between quantum curves and BPS counting that we analyze, and in particular integrality of BPS invariants associated to a given quantum curve, provides an interesting perspective on these problems. An important aspect of our work is an explicit computation of dual extremal quantum A-polynomials for some twist and torus knots, summarized in (\ref{calAxyP-1}) and (\ref{A-torusknots}) and in the attached Mathematica file. In particular an interesting toy model of quantum BPS invariants arises as $m=2$ case of (\ref{calAxyP-1}), which defines a novel $q$-deformed version of Catalan numbers that encode integral invariants; analogous results for other values of $m$ define interesting $q$-deformations of Fuss-Catalan numbers.

Let us stress that one of the motivations for this work have been the results of Markus Reineke on Donaldson-Thomas invariants for $m$-loop quivers \cite{Rei12}. It turns out that these particular invariants are closely related to extremal LMOV invariants for framed unknot and twist knots. In general combinatorial models presented in this work are motivated by the construction in \cite{Rei12}, and after some redefinitions reduce to that construction in case of framed unknot or twist knots. For this reason some of our notation follows \cite{Rei12} and we discuss relations to that work when appropriate. In particular the results of \cite{Rei12} imply that all maximal LMOV invariants for framed unknot and twist knots are integer, which immediately proves integrality of corresponding classical LMOV invariants for twist knots and divisibility statements, discussed in \cite{Garoufalidis:2015ewa}. What is novel in our approach is that we associate combinatorial models to quantum curves (which have not been discussed in the context of Donaldson-Thomas invariants for quivers), our construction works for quite general class of quantum curves (not restricted to a rather special class of difference equations related to $m$-loop quivers), and it leads to interesting results in the realm of knot invariants, seemingly unrelated to \cite{Rei12}. 

The main results of this work are as follows. First, we introduce a generating function of unnormalized colored (extremal) HOMFLY polynomials 
\be
P(x,q) = \sum_r P_r(q) x^r =  \prod_{r\geq 1;j;l\geq 0} \Big(1 - x^r q^{j+2l+1} \Big)^{N_{r,j}}    \label{Pxq-LMOV}
\ee
whose product decomposition that involves LMOV invariants $N_{r,j}$ in exponents follows from the general LMOV decomposition \cite{Labastida:2000zp,Labastida:2000yw}. It can also be shown \cite{Garoufalidis:2015ewa} that $P(x,q)$ satisfies a difference equation that can be written in the form
\be
\widehat{\mathcal{A}}(\hat x, \hat y,q) P(x,q) = 0,  \label{Ahat-intro}
\ee
where $\widehat{\mathcal{A}}(\hat x, \hat y,q)$ is an (extremal) dual quantum A-polynomial (which is simply related to the operator that encodes recursion relations for colored polynomials $P_r(q)$), $\hat x$ acts by multiplication by $x$, and $\hat y P(x,q)=P(qx,q)$. We then argue that, instead of considering colored polynomials $P_r(q)$ or their generating series $P(x,q)$, it is of advantage to focus on the ratio $Y(x,q) = \frac{P(q^2x,q)}{P(x,q)}$, which can be regarded as a functional representation of the operator $\hat y^2$.

Our main result is a construction of a combinatorial model, whose building blocks are encoded in coefficients of the (dual) quantum A-polynomial $\widehat{\mathcal{A}}(\hat x, \hat y,q)$ and can be interpreted as letters in a formal language. One can build words and sentences (series of words) out of these letters. There are two gradings in this model: each letter has a weight $q$ and each word (created out of original letters) in a given sentence is weighted by $x$. This model is designed in such a way that its generating function (Hilbert-Poincar{\'e} series) reproduces $Y(x,q)$
\be
Y(x,q) = \frac{P(q^2x,q)}{P(x,q)} = \sum_{n=0}^{\infty} Y_n(q) x^n = \sum_{n=0}^{\infty}  \Big(\sum_{s\in T_{n}} \textrm{sgn}(s)q^{\textrm{wt}(s)}\Big)x^{n},    \label{Yxq-intro}
\ee
where $\textrm{sgn}(s)$ denotes a sign assigned to a sentence $s$, $\textrm{wt}(s)$ denotes the total number of original letters in a given sentence, and $T_n$ is a (finite) set of sentences consisting of $n$ words and built recursively according to the rules that we specify in detail in what follows. In general, we believe that combinatorial properties of coefficients $Y_n(q)$ deserve thorough studies, especially in the context of knot theory.

A further motivation to construct the combinatorial model is that, apart from reproducing $Y(x,q)$ according to (\ref{Yxq-intro}), it provides insight into the structure of LMOV invariants. Namely, regarding sentences built out of original letters as words in a new language, one can consider a set $T^L$ of Lyndon words in this language. A Lyndon word, defined as a word that is lexicographically strictly smaller than all its cyclic shifts, is one of basic notions in the field known as combinatorics on words \cite{Lothaire1983,Lothaire2002,Berstel2007996}. In order to take into account signs that appear in the decomposition (\ref{Pxq-LMOV}) we enlarge slightly a set of Lyndon words and construct related sets $T^{L,+}_r$ consisting of sentences of length $r$, such that BPS numbers are reconstructed as
\be
\sum_j N_{r,j}q^{j+1} = \frac{1}{[r]_{q^{2}}}\sum_{s\in T_{r}^{L,+}} \textrm{sgn}(s)q^{\textrm{wt}(s)},   \label{Nrj-intro}
\ee
where $[r]_{q^{2}}=\frac{1-q^{2r}}{1-q^2}$ is a standard $q^2$-number. The integrality of $N_{r,j}$ requires that the sum on the right hand side of the above equation is divisible by $[r]_{q^{2}}$, which is a non-trivial condition that can be regarded as a reformulation and sharpening of the LMOV conjecture. For framed unknot and twist knots such divisibility follows from the results in \cite{Rei12}, and we also verify it for some range of $r$ for various torus knots. 

The combinatorial model that we construct leads to other interesting results. First, we deduce from it recursion relations directly for LMOV invariants $N_{r,j}$. Second, in the classical limit $q\to 1$ the dual quantum A-polynomial (\ref{Ahat-intro}) reduces to a classical algebraic curve referred to as a dual extremal A-polynomial in \cite{Garoufalidis:2015ewa}
\be
\mathcal{A}(x,y) = 0,
\ee
whose solution $y=y(x)$ decomposes as
\be
y(x)^2 = Y(x,1) = \prod_{r=1}^{\infty} \big(1 - x^r \big)^{-r b_r} 
\ee
and encodes classical LMOV invariants $b_r=\sum_j N_{r,j}$. In terms of the combinatorial model
\be
b_r = \sum_j N_{r,j} = \frac{1}{r} \sum_{s\in T_{r}^{L,+}} \textrm{sgn}(s),
\ee
so the integrality condition for classical LMOV invariants amounts to the statement that for each $r$ the sum in the above expression is divisible by $r$. The interplay between classical LMOV invariants and algebraic curves was analyzed in \cite{Garoufalidis:2015ewa}, and the above statements explain how those results are related to combinatorial models discussed here. 

The results presented in this paper could be generalized in various directions. It is desirable to prove divisibility by $[r]_{q^2}$ in (\ref{Nrj-intro}) for all $r$, and hence integrality of all extremal LMOV invariants, for other classes of knots. Such relations should be interesting also from the viewpoint of number theory, similarly as discussed in \cite{Garoufalidis:2015ewa}. Apart from extremal invariants, it should be intersting to consider full colored HOMFLY polynomials and include dependence on $a$ in combinatorial models that we construct. Similarly a dependence on the Poincar{\'e} parameter $t$ could be included, and models that we consider could be related to colored homological invariants (knot homologies, superpolynomials, super-A-polynomials), considered e.g. in \cite{Gukov:2011ry,superA,FGSS}. In general, combinatorial interpretation of $Y_n(q)$ introduced in (\ref{Yxq-intro}) deserves further studies and might lead to interesting reformulations of standard knot invariants. Furthermore, relations between BPS invariants and quantum A-polynomials that we discuss should shed light on quantization of algebraic curves  \cite{abmodel,Norbury-quantum,Schwarz:2014hfa,Dumitrescu:2015mpa,Manabe:2015kbj,Bouchard:2016obz}.

The plan of this paper is as follows. In section \ref{sec-review} we review a construction of extremal LMOV invariants and dual A-polynomials. In section \ref{sub:Recursion-relation-induced} we present a construction of a combinatorial model for BPS states and discuss its relations to quantum A-polynomials and combinatorics on words. In section \ref{sec-examples} we illustrate our results in examples that include twist and torus knots, and peculiar $q$-deformations of Catalan numbers. In the appendix we present some technical computations, discuss relations to results in \cite{Rei12}, and provide explicit form of LMOV invariants in various examples.


\section{(Extremal) BPS invariants and (dual) A-polynomials}     \label{sec-review}


In this section we recall two important features of HOMFLY polynomials colored by symmetric representations: on one hand they encode Labastida-Mari{\~n}o-Ooguri-Vafa (LMOV) invariants, and on the other hand they satisfy recursion relations, which can be encoded in quantum A-polynomials. We also introduce corresponding extremal invariants, following \cite{Garoufalidis:2015ewa}.

First we recall the construction of LMOV invariants \cite{OoguriV,Labastida:2000zp,Labastida:2000yw} and present its specialization to the case of $S^r$-colored and extremal HOMFLY polynomials \cite{Garoufalidis:2015ewa}. The starting point is to consider the Ooguri-Vafa generating function 
\be
Z(U,V) = \sum_R  \textrm{Tr}_R U \, \textrm{Tr}_R V = \exp\Big(  \sum_{n=1}^{\infty} \frac{1}{n} \Tr U^n \Tr V^n \Big),
\ee
where $U=P\,\exp\oint_K A$ is the holonomy of $U(N)$ Chern-Simons gauge field along a knot $K$, $V$ can be interpreted as a source, and the sum runs over all representations $R$, i.e. all two-dimensional partitions. The LMOV conjecture states that 
\be
\big\langle Z(U,V) \big\rangle = \sum_R P_{R}(a,q) \textrm{Tr}_R V  = \exp \Big(  \sum_{n=1}^\infty \sum_R \frac{1}{n} f_{R}(a^n,q^n) \textrm{Tr}_R V^n  \Big),    \label{ZUV}
\ee
where the expectation value of the holonomy is identified with the unreduced HOMFLY polynomial of a knot $K$, $\langle \textrm{Tr}_R U \rangle = P_{R}(a,q)$,  for the unknot in the fundamental representation normalized as $P^{{\bf 0_1}}_{\square}(a,q)=\frac{a-a^{-1}}{q-q^{-1}}$. The functions $f_{R}(a,q)$ take form
\be
f_{R}(a,q) = \sum_{i,j} \frac{N_{R,i,j} a^i q^j}{q-q^{-1}},  \label{fR}
\ee
where $N_{R,i,j}$ are conjecturally integer BPS degeneracies (LMOV invariants), which count M2-branes ending on M5-branes that wrap a Lagrangian submanifold associated to a given knot $K$ in the conifold geometry. 
For a fixed $R$ there is a finite number of non-zero $N_{R,i,j}$. 

Consider now a one-dimensional $V \equiv x$. In this case $\Tr_R V \neq 0$ only for symmetric representations $R=S^r$ (labeled by partitions with a single row with $r$ boxes) and $\textrm{Tr}_{S^r}(x) = x^r$. 
Denoting $P_r(a,q) = P_{S^r}(s,q)$, $N_{r,i,j}=N_{S^r,i,j}$, $f_r=f_{S^r}$, in this case we can write (\ref{ZUV}) as
\be
P(x,a,q) = \sum_{r=0}^\infty P_{r}(a,q) x^r = \exp\Big( \sum_{r,n\geq 1} \frac{1}{n} f_{r}(a^n,q^n)x^{n r}\Big) = \prod_{r\geq 1;i,j;l\geq 0} \Big(1 - x^r a^i q^{j+2l+1} \Big)^{N_{r,i,j}},
\label{Pz2}
\ee
so that $f_r(a,q)$ are expressed solely in terms of $S^r$-colored HOMFLY polynomials, e.g.
\bea
f_1(a,q) &=& P_1(a,q), \qquad
f_2(a,q) = P_2(a,q) - \frac{1}{2}P_1(a,q)^2 -\frac{1}{2} P_1(a^2,q^2), \\
f_3(a,q) &=& P_3(a,q) - P_1(a,q)P_2(a,q) + \frac{1}{3}P_1(a,q)^3 - \frac{1}{3} P_1(a^3,q^3). \label{f-P}  
\eea
In consequence LMOV invariants $N_{r,i,j}$ for symmetric representations can be consistently defined and form a closed system.



Furthermore, we recall that $S^r$-colored HOMFLY polynomials satisfy a linear $q$-difference equation \cite{Garoufalidis:2016zhf}, which for all $r\in \mathbb{Z}$ (with $P_r(a,q)=0$ for $r<0$) can be written in terms of an operator $\widehat{A}(\widehat{M},\widehat{L},a,q)$ called quantum ($a$-deformed) A-polynomial 
\begin{equation}
\widehat{A}(\widehat{M},\widehat{L},a,q)P_{r}(a,q) \equiv \left(\sum_{m,l} A_{l,m}(a,q)\widehat{M}^{2m}\widehat{L}^{l}\right) P_r(a,q) = 0,   \label{Ahat-a}
\end{equation}
where $\widehat{M}$ and $\hat{L}$ are operators that satisfy the relation $\widehat{M}\widehat{L}=q\widehat{L}\widehat{M}$ and are represented as
\begin{equation}
\widehat{M}P_{r}(a,q)=q^{r}P_{r}(a,q),\qquad \widehat{L}P_{r}(a,q)=P_{r+1}(a,q).\label{MhatLhat}
\end{equation}
Multiplying (\ref{Ahat-a}) by $x^{r}$, summing over all integers $r$, then acting with $\widehat{L}^{l}\widehat{M}^{2m}$, and denoting the maximal power of $\widehat{L}$ by $l^{max}$, we  can transform (\ref{Ahat-a}) into 
\begin{equation}
\sum_{l,m}A_{l,m}(a,q)\hat{x}^{l^{max}-l}\hat{y}^{2m}P(x,a,q)=0,   \label{eq:multiplied by lmax}
\end{equation}
where $\hat x$ and $\hat y$ are operators acting on the generating function $P(x,a,q)$ as
\begin{equation}
\widehat{x}P(x,a,q)=xP(x,a,q),\qquad \widehat{y}P(x,a,q)=P(qx,a,q).\label{eq:x & y def}
\end{equation}
Finally, we define a dual quantum A-polynomial
\be
\widehat{\mathcal{A}}(\hat{x},\hat{y},a,q)=\sum_{l,m}\mathcal{A}_{l,m}(a,q)\hat{x}^{l}\hat{y}^{2m}, \qquad \mathcal{A}_{l,m}(a,q)=A_{l^{max}-l,m}(a,q),
\ee
in terms of which (\ref{eq:multiplied by lmax}) is written simply as 
\begin{equation}
\widehat{\mathcal{A}}(\hat{x},\hat{y},a,q)P(x,a,q)=0.\label{calAxyP}
\end{equation}

In the limit $q\to 1$ one can consider classical versions of (dual) A-polynomials and LMOV invariants. In this limit the dual quantum A-polynomial reduces to an algebraic curve
\be
\mathcal{A}(x,y,a) = 0,
\ee
whose solution $y=y(x)=\lim_{q\to 1} \frac{P(qx)}{P(x)}$ decomposes as
\be
y(x) = \prod_{r=1}^{\infty} \big(1 - x^r a^i \big)^{-r b_{r,i}/2}     \label{yx-bri}
\ee
where $b_{r,i} = \sum_j N_{r,i,j}$ are classical LMOV invariants.



Following \cite{Garoufalidis:2015ewa}, we can also restrict the results reviewed above to extremal cases, i.e. focus only on coefficients of lowest or highest (bottom and top) powers of $a$ in colored HOMFLY polynomials. To this end we focus on (a large class of) knots that satisfy
\be
P_r(a,q) = \sum_{i=r \cdot c_-}^{r\cdot c_+} a^i p_{r,i}(q)      
\label{Pr-minmax}
\ee
for some integers $c_\pm$ and for every natural number $r$, where $p_{r,r \cdot c_\pm}(q) \neq 0$, and define
\be
P^\pm_r(q) = p_{r,r \cdot c_\pm}(q).         \label{Pminmax def}
\ee
Likewise, we can consistently introduce extremal LMOV invariants $N^\pm_{r,j} = N_{r, r \cdot c_\pm,j}$, so that
\be
P^{\pm}(x,q) \equiv \sum_r P^\pm_r(q) x ^r = \prod_{r\geq 1;j;l\geq 0} \Big(1 - x^r q^{j+2l+1} \Big)^{N^\pm_{r,j}} .         \label{Pminmax product form}
\ee
If $P_r(a,q)$ is annihilated by $\widehat{A}(\hat M, \hat L,a,q)$, then its extremal part $P^\pm_r(q)$ is annihilated by the operator $\widehat{A}^\pm(\hat M, \hat L,q)$ obtained by multiplying 
$\widehat{A}(\hat M, \hat L,a^{\mp 1},q)$ by $a^{\pm r c_\pm}$ 
and then setting $a=0$, so that (\ref{calAxyP}) reduces to 
\be
\widehat{\mathcal{A}}^{\pm}(\hat{x},\hat{y},q)P^{\pm}(x,q) \equiv \Big(  \sum_{l,m}\mathcal{A}^{\pm}_{l,m}(q)\hat{x}^{l}\hat{y}^{2m} \Big) P^{\pm}(x,q) = 0. \label{extremal calAxyP}
\ee
In the classical limit we obtain extremal dual A-polynomial equation $\mathcal{A}^{\pm}(x,y)=0$, whose solution $y=y(x)$ encodes extremal classical LMOV invariants $b_r^{\pm}$
\be
y=\prod_r(1 - x^r)^{-r b^{\pm}_r/2}, \qquad b_r^{\pm}=\sum_j N^{\pm}_{r,j}.   \label{y-br}
\ee

In most of this paper we focus on extremal invariants, so we often suppress superscripts $\pm$ and denote extremal HOMFLY polynomials and their generating series, LMOV invariants, dual A-polynomials, etc. simply as $P_r(q), P(x,q), N_{r,j}, b_r, \widehat{\mathcal{A}}(\hat{x},\hat{y},q), \mathcal{A}_{l,m}, \mathcal{A}(x,y)$, etc.


\section{BPS counting and combinatorics on words}  \label{sub:Recursion-relation-induced}

In this section we introduce a combinatorial model for BPS state counting. We focus on extremal invariants and suppress indices $\pm$ in various expressions. A generalization to full HOMFLY polynomials (including $a$-dependence) or superpolynomials (depending on an additional parameter $t$) is also possible, however the extremal case enables us to illustrate the essence of the construction, without additional technical complications. 

First, we propose to consider the following ratio of generating functions (\ref{Pminmax product form}), which can be considered as a functional representation of $\hat y^2$ operator 
\begin{equation}
Y(x,q)=\frac{P(q^{2}x,q)}{P(x,q)} = \prod_{r=1}^{\infty} \prod_j \prod_{l=0}^{r-1}   \Big(1-x^{r}q^{j+2l+1}\Big)^{-N_{r,j}} \equiv \sum_{n=0}^{\infty} Y_n(q) x^n,   \label{eq:Y def}
\end{equation}
where coefficients $Y_n(q)$ on the right hand side are defined upon an expansion in $x$ and in particular $Y_0(q)=1$. The function $Y(x,q)$, similarly to $P(x,q)$ in (\ref{Pminmax product form}), encodes all quantum LMOV invariants $N_{r,j}$, however it has an important advantage: $Y_n(q)$, as a coefficient at $x^n$, is a finite polynomial in $q$ (this is not so in case of (\ref{Pminmax product form}), for which coefficients at various powers of $x$ are rational functions in $q$). This is a crucial feature that enables to construct a combinatorial model. In the classical limit $q\to 1$, $Y(x,1)$ is identified as a square of (\ref{yx-bri}) that solves the classical dual extremal A-polynomial equation $\mathcal{A}(x,y)=0$.

Note that dividing (\ref{extremal calAxyP}) by $P(x,q)$, it can be rewritten in the form
\begin{equation}
\mathcal{A}(x,Y(x,q),q) \equiv \sum_{l,m}\mathcal{A}_{l,m}(q)x^{l}Y(x,q)^{(m;q^{2})}=0,      \label{eq:A(x,Y(x,q),q)}
\end{equation}
where the $m$'th $q$-power of a function $f(x,q)$ is defined as
\begin{equation}
f(x,q)^{(m;q)} = \prod_{i=0}^{m-1}    f(q^{i}x,q).
\end{equation}
Our construction of the combinatorial model will be based on the recursive analysis of the equation (\ref{eq:A(x,Y(x,q),q)}), which is expressed in terms of coefficients in the extremal A-polynomial $\mathcal{A}_{l,m}(q)$. It is clear the these coefficients cannot be arbitrary -- the existence of integer LMOV invariants imposes strong constraints on the form of the generating function (\ref{Pminmax product form}), and so on the equation it satisfies. While precise conditions on A-polynomials that guarantee integrality are quite subtle \cite{abmodel}, in what follows we consider a large class of equations (\ref{eq:A(x,Y(x,q),q)}) of the form
\begin{equation}
\mathcal{A}(x,Y(x,q),q) = 1-Y(x,q)+\sum_{l\geq1,m\geq1}\mathcal{A}_{l,m}(q)x^{l}Y(x,q)^{(m;q^{2})} +\sum_{l\in\Lambda}\mathcal{A}_{l,0}(q)x^{l} = 0,     \label{eq:nonhom-Y}
\end{equation}
where $\Lambda$ is finite subset of $\mathbb{N}$. In the above equation coefficients $\mathcal{A}_{l,m}(q)$ take form
\be
\mathcal{A}_{0,0}(q)=1,\qquad  \mathcal{A}_{0,1}(q)=-1,\qquad  \mathcal{A}_{0,m}(q)=0\ \forall m\geq2.
\ee
All examples of quantum A-polynomials for knots that we analyzed are of this form. In particular, this form implies that $Y(x,q)\sim 1$ for small $x$, which is consistent with (\ref{eq:Y def}).

Due to the presence of the last term $\sum_{l\in\Lambda}\mathcal{A}_{l,0}(q)x^{l}$ we call the equation (\ref{eq:nonhom-Y}) as nonhomogeneous. In what follows we consider first a homogeneous equation
\begin{equation}
\mathcal{A}(x,Y(x,q),q)=1-Y(x,q)+\underset{l\geq1,m\geq1}{\sum}\mathcal{A}_{l,m}(q)x^{l}Y(x,q)^{(m;q^{2})}=0,   \label{eq:Equation for Y in good class for recursion}
\end{equation}
which is characterized by $\mathcal{A}_{l,0}(q)=0$ for $l\geq 1$. The combinatorial model associated to the nonhomogeneous equation (\ref{eq:nonhom-Y}) is a generalization of the model for the homogeneous case, and in fact, depending on a knot, A-polynomials yield either homogeneous or nonhomogeneous equations, so in any case it is important to analyze both these cases. For this reason we present first a construction of the combinatorial model for the case of (\ref{eq:Equation for Y in good class for recursion}), and subsequently generalize it to the case of (\ref{eq:nonhom-Y}). In what follows we also use the notation
\begin{equation}
\mathcal{A}_{l,m}(q)=\sum_j \mathcal{A}_{l,m,j}q^{j}.
\end{equation}

Our aim is to construct a combinatorial model associated to $\widehat{\mathcal{A}}(\hat x, \hat y,q)$, whose generating function is equal to $Y(x,q)$ 
\begin{equation}
Y(x,q)=\sum_{n=0}^{\infty}  Y_{n}(q)x^{n} = \sum_{n=0}^{\infty} \left(\sum_{s\in T_{n}} \textrm{sgn}(s)q^{\textrm{wt}(s)}\right)x^{n}.     \label{eq:Combinatorial generation of Y}
\end{equation}
Therefore $Y(x,q)$ can be thought of as a (signed) Hilbert-Poincar{\'e} series of a bigraded free algebra $B$ whose basis is a graded set $T= \bigcup_{n=0}^{\infty} T_{n}$, $\textrm{sgn}(s)$ denotes a sign of an element $s$, and an integer-valued weight $\textrm{wt}(s)$ provides the second grading \cite{Rei12}.


\subsection{Combinatorial model, homogeneous case}  \label{sub:Combinatorial-construction for coefficients of Y}

We construct first a combinatorial model associated to a homogeneous equation (\ref{eq:Equation for Y in good class for recursion}). Note that expanding (\ref{eq:Equation for Y in good class for recursion}) in powers of $x$ we obtain recursion relations for $Y_n(q)$ introduced in (\ref{eq:Y def})
\begin{equation}
Y_{n}(q)=\sum_{l\geq1,m\geq1}\mathcal{A}_{l,m}(q)\sum_{k_{0}+\ldots+k_{m-1}=n-l}\Big(\prod_{i=0}^{m-1} q^{2ik_{i}}Y_{k_{i}}(q)\Big),    \label{eq:Recursion}
\end{equation}
with the initial condition $Y_0(q)=1$. Our first aim is to construct the set $T= \bigcup_{n=0}^{\infty} T_{n}$ introduced in (\ref{eq:Combinatorial generation of Y}) in a way consistent with the recursion (\ref{eq:Recursion}), which in particular suggests that $T_{n}$ should be obtained by concatenation of elements of $T_{k_{0}}, T_{k_{1}}, \ldots, T_{k_{m-1}}$. 

Our construction of $T$ is based on the notion of a formal language, natural in the context of a free algebra. We recall first a few basic definitions. Consider a countable, totally ordered set $\Sigma$ called an \emph{alphabet}, whose elements are \emph{letters}. Strings of letters are called \emph{words}; an empty word is denoted $\varepsilon$. The set of all words made of letters from the alphabet $\Sigma$ is denoted by $\Sigma^{*}$. Lists of words are called \emph{sentences}. We denote by $\Sigma^{**}$ the set of all sentences made of words from $\Sigma^{*}$. For appropriately defined alphabet $\Sigma$, our set $T$ will arise as a subset of $\Sigma^{**}$.

The length $\textrm{lt}(s)$ of a sentence $s$ is defined as the number of words it consists of. The weight $\textrm{wt}(\mu)$ of a word $\mu$ is defined as the number of letters it consists of. We also define an antiword $\bar{\mu}$ as a word $\mu$ with the opposite weight assigned,  $\textrm{wt}(\bar{\mu})=-\textrm{wt}(\mu)$. A weight of a sentence $s=\left[\sigma_{1},\ldots,\sigma_{S}\right]$ is defined by $\textrm{wt}(s)=\sum_{i=1}^S \textrm{wt}(\sigma_{i})$. Note that $\textrm{wt}([\,])=\textrm{wt}([\varepsilon])=0$, so $\textrm{wt}(s)$ is insensitive to the number of
words in $s$ (as $[\,]$ contains no words whereas $[\varepsilon]$ contains one). We denote a concatenation of two words $\mu,\nu\in\Sigma^{*}$ by $\mu*\nu$, and for positive $j$ we define
\begin{equation}
\mu^{*j}=\underset{j}{\underbrace{\mu*\mu*\cdots*\mu}},\qquad    \mu^{*\left(-j\right)}=\underset{j}{\underbrace{\bar{\mu}*\bar{\mu}*\cdots*\bar{\mu}}.}    \label{eq:words concatenation power}
\end{equation}
Concatenation of sentences $s=\left[\sigma_{1},\ldots,\sigma_{S}\right]$ and $t=\left[\tau_{1},\ldots,\tau_{T}\right]$ is also denoted by $*$
\begin{equation}
s*t = \left[\sigma_{1},\ldots,\sigma_{S},\tau_{1},\ldots,\tau_{T}\right], \qquad   s^{*j} = \underset{j}{\underbrace{s*s*\cdots*s}}.   \label{eq:concatenation of sentences}
\end{equation}
In particular for a sentence consisting of one word $[\mu]$ 
\begin{equation}
[\mu]^{*j}=\underset{j}{\underbrace{[\mu]*[\mu]*\cdots*[\mu]}}=\underset{j}{\underbrace{\left[\mu,\mu,\cdots,\mu\right]}.}
\end{equation}
For two sentences of the same length $s=[\sigma_{1},\ldots,\sigma_{S}]$ and  $t=[\tau_{1},\ldots,\tau_{S}]$ we also define
\begin{equation}
s\vee t = [\sigma_{1}*\tau_{1},\ldots,\sigma_{S}*\tau_{S}].
\end{equation}

We can present now a recursive construction of a combinatorial model. The initial condition $Y_{0}(q)=1$ means that $T_{0}$ consists of one element of trivial weight, so that it is natural to identify it with the empty list
\begin{equation}
T_{0} = \{[\,]\}.   \label{eq:T_0}
\end{equation}
Furthermore we choose an alphabet $\Sigma$ that consists of $I=\sum_{l\geq1,m\geq1,j}\left|\mathcal{A}_{l,m,j}\right|$ letters and out of those letters construct $I$ different one letter words, which we assign uniquely to all $I$ units represented by coefficients in the relation (\ref{eq:Recursion}). In order to define the recursion step that determines $T_{n}$ let us:
\begin{itemize}
\item assume that we have constructed sets $T_{0}, T_{1},\dots, T_{n-1}$,
\item fix a partition $k_{0}+\ldots+k_{m-1}=n-l$ (\emph{without} demanding $k_{0}\leq\ldots\leq k_{m-1}$) together with $m$ sentences $s_{k_{0}}, s_{k_{1}},\ldots,s_{k_{m-1}}$ from $T_{k_{0}}, T_{k_{1}},\ldots,T_{k_{m-1}}$
respectively,
\item fix $l$, $m$, $j$ for which $\mathcal{A}_{l,m,j}$ is non-vanishing, 
\item fix a one letter word $\mu$ corresponding to one unit in $\mathcal{A}_{l,m,j}$. 
\end{itemize}
Then we define a new sentence
\begin{equation}
s(l,m,j;s_{k_{0}},\ldots,s_{k_{m-1}};\mu)=\label{eq:new sentence}
\end{equation}
\[
=\textrm{sgn}\left(\mathcal{A}_{l,m,j}\right)\left[\varepsilon\right]{}^{*\left(l-1\right)}*[\mu^{*j}]*\left([\mu^{*2\left(m-1\right)}]{}^{*k_{m-1}}\vee s_{k_{m-1}}\right)*\cdots*\left([\mu^{*2}]{}^{*k_{1}}\vee s_{k_{1}}\right)*s_{k_{0}}
\]
where $\varepsilon$ is an empty word. As sign behaves under concatenation like under multiplication,
\begin{equation}
\textrm{sgn}\left(s(l,m,j;s_{k_{0}},\ldots,s_{k_{m-1}};\mu)\right)=\textrm{sgn}\left(\mathcal{A}_{l,m,j}\right)\underset{i=0}{\overset{m-1}{\prod}}\textrm{sgn}\left(s_{k_{i}}\right).
\end{equation}
We define $T_{n}$ as a set of all sentences $s$ constructed in (\ref{eq:new sentence}), considering all possible choices of $(l,m,j;s_{k_{0}},\ldots,s_{k_{m-1}};\mu)$. Each $T_n$ consists therefore of sentences of length $n$ and we denote
\begin{equation}
T= \bigcup_{n=0}^{\infty}T_{n}.
\end{equation}
It follows from the above construction that $Y(x,q)$ defined in (\ref{eq:Y def}) can be represented as
\begin{equation}
Y(x,q) = \sum_{n=0}^{\infty}\sum_{s\in T_{n}}  \textrm{sgn}(s) q^{\textrm{wt}(s)}x^{n}=\sum_{s\in T}\textrm{sgn}(s)q^{\textrm{wt}(s)}x^{\textrm{lt}(s)},   \label{Yxq-Tn}
\end{equation} 
which is the result (\ref{eq:Combinatorial generation of Y}) that we have been after.

Note that if we define $B$ as the free algebra generated by elements of $T$ with the multiplication identified with the concatenation of sentences (\ref{eq:concatenation of sentences}), and bigraded by the number of words and the number of letters, then its Hilbert-Poincar{\'e} series is equal to
\begin{equation}
HP(B)=\overset{\infty}{\underset{n=0}{\sum}}\overset{\infty}{\underset{j=0}{\sum}} q^{j}x^{n} \textrm{dim}B_{n,j}=\underset{s\in T}{\sum}q^{\textrm{wt}(s)}x^{\textrm{lt}(s)},\label{eq:Hilbert-Poincare series}
\end{equation}
where $B_{n,j}$ is generated by all sentences of $n$ words and $j$ letters (so $\textrm{dim}B_{n,j}$ is the number of such sentences in $T$). Therefore (\ref{Yxq-Tn}) can be regarded as a signed analogue of $HP(B)$.

In what follows we also illustrate the above construction graphically, by representing words as columns of labeled boxes with letters (growing upwards), and sentences as horizontal series of columns. Therefore elements of $T_{n}$ consist of $n$ columns and their weight is given by the total number of boxes in all those columns (excluding boxes with an empty word $\varepsilon$). 
Here is an example of a sentence made of 3 words and of weight 5:
\begin{equation}
\left[\mu,\mu\mu\nu,\nu\right]\sim\young(:\nu:,:\mu:,\mu\mu\nu).
\end{equation}


\subsection{Extremal LMOV invariants from Lyndon words} \label{sub:Combinatorial-construction-BPS}

Having expressed $Y(x,q)$ as a generating series of the combinatorial model described above, we now show that LMOV invariants encoded in (\ref{eq:Y def}) also have a natural interpretation in this model and are related to an important notion of Lyndon words. In what follows we consider the following combinations of LMOV invariants $N_{r,j}$
\begin{equation}
N_{r}(q) = \underset{j}{\sum}N_{r,j}q^{j+1}.  \label{Nrq-def}
\end{equation}

To start with we introduce a set $T^{0}\subset T$ of \emph{primary sentences}, i.e. sentences which cannot be presented as a concatenation of other sentences
\begin{equation}
T^{0} = \left\{ s\in T:\ s\neq s_{1}*s_{2}*\ldots*s_{t}\ \forall s_{i}\in T,\ t>1\right\}.     \label{eq:T zero definition}
\end{equation}
This set decomposes into subsets of primary sentences of length $n$
\begin{equation}
T_{n}^{0}=T^{0}\cap T_{n}.   \label{eq:Tzero product}
\end{equation}
Elements of $T^{0}$ generate a free algebra, which we denote by $B^{0}$. Since every sentence from $T$ can be uniquely represented as a concatenation of primary sentences, there is an isomorhpism of a tensor algebra $T\left(B^{0}\right)$ and the algebra $B$ 
\begin{equation}
T\left(B^{0}\right)\cong B.   \label{eq:BTB0}
\end{equation}
This isomorphism induces a bijection
\begin{equation}
\left(T^{0}\right)^{*}\overset{\varphi}{\longrightarrow}T,
\end{equation}
where $\left(T^{0}\right)^{*}$ is a formal language over an alphabet $T^{0}$ with a lexicographic ordering induced by one from $\Sigma^{**}$.
The isomorphism $\varphi$ on $s\in T^{0}$ is defined as
\begin{equation}
\varphi:\ s=w\mapsto s=\left[\sigma_{1},\ldots,\sigma_{S}\right],\label{eq:phidef}
\end{equation}
where on the left hand side we treat $s$ as a one-letter word $w\in\left(T^{0}\right)^{*}$,
whereas on the right hand side $s$ is a sentence $\left[\sigma_{1},\ldots,\sigma_{S}\right]\in\Sigma^{**}$. The action of $\varphi$ on words that contain more letters can be obtained from the fact that $\varphi$ translates concatenation of words in $\left(T^{0}\right)^{*}$ into concatenation of sentences in $\Sigma^{**}$
\begin{equation}
\varphi\left(w_{1}*w_{2}\right)=\varphi\left(w_{1}\right)*\varphi\left(w_{2}\right).    \label{eq:phiconc}
\end{equation}
Note that the notion of words has now a multiple meaning, which we hope will be clear from the context. In particular words in the language $\left(T^{0}\right)^{*}$ can be identified with elements of $T$, which are sentences from $\Sigma^{**}$. 

Let us recall now a definition of a Lyndon word: it is a word that is lexicographically strictly smaller than all its cyclic shifts. For example, in the usual lexicographic ordering $[abcd]$ is a Lyndon word, because it is smaller than all its cyclic shifts $[bcda]$, $[cdab]$, and $[dabc]$. An important Chen-Fox-Lyndon theorem asserts that every word can be written in a unique way as a concatenation of Lyndon words, weakly decreasing lexicographically \cite{Lothaire1983,Lothaire2002}. 

Consider now a set of all Lyndon words in the language $\left(T^{0}\right)^{*}$ and denote by $T^{L}$ the image of this set under $\varphi$. $T^{L}$ is doubly graded by the number of words and the number of letters and in analogy to (\ref{eq:Tzero product}) can be decomposed into subsets of length $n$
\begin{equation}
T_{n}^{L} = T^{L}\cap T_{n}.    \label{eq:TL product-1}
\end{equation}

Let us rewrite now the generating series (\ref{Yxq-Tn}) taking advantage of the Chen-Fox-Lyndon theorem. Consider first the Hilbert-Poincar{\'e} series (\ref{eq:Hilbert-Poincare series}) and note, that the Chen-Fox-Lyndon theorem implies that every term in the expression $\sum_{s\in T\cong\left(T^{0}\right)^{*}}q^{\textrm{wt}(s)}x^{\textrm{lt}(s)}$ corresponds to a product of Lyndon words. Since in the Chen-Fox-Lyndon theorem factors decrease weakly, we have to consider all possible numbers of copies $s^{*i}$ of a given Lyndon word $s$. Ordinary multiplication is commutative so the order in the product over Lyndon words does not matter, although keeping it fixed is crucial for proper counting. It follows that the Hilbert-Poincar{\'e} series (\ref{eq:Hilbert-Poincare series}) can be determined by considering the product of generators corresponding to Lyndon words
\begin{equation}
HP(B)=\prod_{s\in T^{L}} \Big(\sum_{i=0}^{\infty} q^{\textrm{wt}(s^{*i})}x^{\textrm{lt}(s^{*i})}\Big)
=\prod_{s\in T^{L}} \Big(\sum_{i=0}^{\infty}\big(q^{\textrm{wt}(s)}x^{\textrm{lt}(s)}\big)^{i}\Big)
=\prod_{s\in T^{L}} \frac{1}{1-q^{\textrm{wt}(s)}x^{\textrm{lt}(s)}}.    \label{HPB-prod}
\end{equation}
Analogously, the generating series $Y(x,q)$ can be written as
\begin{equation}\label{eq:Y in terms of Lyndon words sign inside bracket}
Y(x,q)= \prod_{s\in T^{L}} \Big(\sum_{i=0}^{\infty}\textrm{sgn}\left(s^{*i}\right)q^{\textrm{wt}(s^{*i})}x^{\textrm{lt}(s^{*i})}\Big)
=\prod_{s\in T^{L}}\frac{1}{1-\textrm{sgn}\left(s\right)q^{\textrm{wt}(s)}x^{\textrm{lt}(s)}},
\end{equation}
and to determine it we have to include the sign dependence in (\ref{HPB-prod}) and change every $q^{\textrm{wt}(s^{*i})}x^{\textrm{lt}(s^{*i})}$ into $\textrm{sgn}\left(s^{*i}\right)q^{\textrm{wt}(s^{*i})}x^{\textrm{lt}(s^{*i})}$. This can be achieved using $1-q^{j}x^{r}=\frac{1-\left(q^{j}x^{r}\right)^{2}}{1+q^{j}x^{r}}$,  
\begin{equation}\label{eq:Y in terms of Lyndon words long version}
Y(x,q)=\Big(\prod_{s\in T^{L},\textrm{sgn}\left(s\right)>0} \frac{1}{1-q^{\textrm{wt}(s)}x^{\textrm{lt}(s)}}   \Big)
\Big( \prod_{s\in T^{L},\textrm{sgn}\left(s\right)<0}\frac{1-q^{\textrm{wt}(s)}x^{\textrm{lt}(s)}}{1-q^{2\textrm{wt}(s)}x^{2\textrm{lt}(s)}}\Big).
\end{equation}
Furthermore, we can treat $\left(1-q^{2\textrm{wt}(s)}x^{2\textrm{lt}(s)}\right)^{-1}$ as coming from extra sentences. Following \cite{Rei12} we define a new set
\begin{equation}
T^{L,+}=T^{L}\cup\left\{ s*s:s\in T^{L},\textrm{sgn}\left(s\right)=-1\right\}   \label{eq:T^L,+}
\end{equation}
and denote by $T_{r}^{L,+}$ a subset of $T^{L,+}$ consisting of sentences of $r$ words, so that
\begin{equation}
T^{L,+}=\bigcup_{r=0}^{\infty}T_{r}^{L,+},
\end{equation}
and by $T_{p,r}^{L,+}$ denote a subset of $T_{r}^{L,+}$ containing sentences of $p$ letters. Note that
\begin{equation}
T^{0}\subset T^{L}\subset T^{L,+}\subset T.
\end{equation}
Now we can write
\begin{equation}
Y(x,q)=\prod_{s\in T^{L,+}} \big(1-q^{\textrm{wt}(s)}x^{\textrm{lt}(s)}\big)^{-\textrm{sgn}\left(s\right)}=\prod_{r\geq1}\prod_p (1-q^{p}x^{r} )^{-Q_{r,p}},\label{eq:Y in terms of Lyndon words final}
\end{equation}
where 
\begin{equation}
Q_{r,p}=\sum_{s\in T_{p,r}^{L,+}} \textrm{sgn}\left(s\right)
\end{equation}
is a net number of elements of $T_{p,r}^{L,+}$. We can interpret the equation (\ref{eq:Y in terms of Lyndon words final}) as the correspondence between elements of $T^{L,+}$ and bosonic (for $\textrm{sgn}\left(s\right)=1$) or fermionic (for $\textrm{sgn}\left(s\right)=-1$) generators. 
In addition we define
\begin{equation}
Q_{r}(q)=\sum_{s\in T_{r}^{L,+}} \textrm{sgn}\left(s\right)q^{\textrm{wt}(s)}=\sum_{p=p_{min}}^{p_{max}}Q_{r,p}q^{p}. \label{eq:Qr(q)}
\end{equation}
By comparison of two expressions for $Y(x,q)$ given in (\ref{eq:Y def}) and (\ref{eq:Y in terms of Lyndon words final})
\begin{equation}
Y(x,q) = \prod_{r=1}^{\infty} \prod_j \prod _{l=0}^{r-1} \Big(1-x^{r}q^{j+2l+1}\Big)^{-N_{r,j}}=
\prod_{r=1}^{\infty}\prod_p \left(1-q^{p}x^{r}\right)^{-Q_{r,p}}\label{eq:Y-product two sides}
\end{equation}
we then find
\begin{equation}
N_{r}(q)=\sum_j N_{r,j}q^{j+1}=\frac{Q_{r}(q)}{[r]_{q^{2}}}=\frac{1}{[r]_{q^{2}}}\sum_{s\in T_{r}^{L,+}} \textrm{sgn}\left(s\right)q^{\textrm{wt}(s)}.  \label{eq:Nr(q)}
\end{equation}
This is one of our main results. Note that $Q_{r,p}$ are integer and can be constructed for any equation of the form (\ref{eq:Equation for Y in good class for recursion}) (as well as (\ref{eq:nonhom-Y}), as discussed in the next section). However divisibility of $Q_{r}(q)$ by $[r]_{q^{2}}=\frac{1-q^{2r}}{1-q^{2}}$ is not guaranteed, and it equivalent to integrality of BPS invariants $N_{r,j}$. 


\subsection{Nonhomogeneous case}  \label{sub:Generalisation-to-knots}

In this section we generalize the above construction to the nonhomogeneous case (\ref{eq:nonhom-Y}), with $\mathcal{A}_{0,0}(q)=-\mathcal{A}_{0,1}(q)=1$, $\mathcal{A}_{0,m}(q)=0$ for $m\geq2$, and $\mathcal{A}_{l,0}(q)=\sum_j \mathcal{A}_{l,0,j}q^{j}\neq 0$. This generalization does not affect the form of the recursion (\ref{eq:Recursion})
for $n>l^{max}$, where $l^{max}$ is the largest element in $\Lambda$. However it modifies expressions for $Y_{n}(q)$ for $n\leq l^{max}$, which we can interpret as a new set of initial conditions. It turns out that we can consider first the homogeneous equation (\ref{eq:Equation for Y in good class for recursion}) as in section \ref{sub:Combinatorial-construction for coefficients of Y}, and modify its solution in order to take the nonhomogeneous term into account. Let us denote sets associated to the homogeneous equation, obtained as in sections \ref{sub:Combinatorial-construction for coefficients of Y} and \ref{sub:Combinatorial-construction-BPS}, with an additional a superscript $hom$. These are the sets of sentences of the form (\ref{eq:new sentence}) $T^{hom}=\bigcup_{n=0}^{\infty} T_{n}^{hom}$, primary sentences $T^{hom,0}$, Lyndon words $T^{hom,L}$ and its modified version, $T^{hom,L,+}$; we also identify the language $\left(T^{hom,0}\right)^{*}$ with $T^{hom}$.

Now we construct a combinatorial model for the nonhomogeneous equation (\ref{eq:nonhom-Y}). Its first ingredient is a set $T^{nonh}=\bigcup_{n=0}^{\infty} T^{nonh}_n$, similarly as before determined recursively, and with the same initial condition as in the homogeneous case
\begin{equation}
T_{0}^{nonh}=T_{0}^{hom}=\{[\,]\}.
\end{equation}
Furthermore, we introduce two sets of letters. First, we consider $\sum_{l\geq1,m\geq1,j}\left|\mathcal{A}_{l,m,j}\right|=I$ letters assigned to the coefficients of the homogeneous equation, in the same way as before. Second, we augment the alphabet $\Sigma$ by $\sum_{l\in\Lambda,j}\left|\mathcal{A}_{l,0,j}\right|=J$ new letters, which are lexicographically strictly smaller than letters from the first set, and assign $\left|\mathcal{A}_{l,0,j}\right|$ one letter words to every $\mathcal{A}_{l,0,j}$ in $\mathcal{A}(x,Y,q)$. We denote one letter words corresponding to every unit in $\sum_{l\in\Lambda,j}\left|\mathcal{A}_{l,0,j}\right|$ by $\alpha,\beta,\gamma,\ldots$, and one letter words corresponding to every unit in $\sum_{l\geq1,m\geq0,j}\left|\mathcal{A}_{l,m,j}\right|$ by $\mu, \nu, \xi,\ldots$.
Now for each one letter word $\alpha$ corresponding to one unit in $\mathcal{A}_{l,0,j}$ we define a new sentence of $l$ words and $j$ letters
\begin{equation}
s(l,j,\alpha)=\textrm{sgn}(\mathcal{A}_{l,0,j})\left[\varepsilon\right]{}^{*\left(l-1\right)}*[\alpha^{*j}].\label{eq:new sentence nonhomogeneous case}
\end{equation}
Note that this can be regarded a generalization of (\ref{eq:new sentence}) to the case $m=0$. 

We assume that the sets 
$T_{0}^{nonh}, T_{1}^{nonh},\ldots,T_{n-1}^{nonh}$ have been already constructed, and define $T_{n}^{nonh}$ recursively. Let us denote by $T_{n}^{aux1}$ the set of all sentences $s(n,j,\alpha)$ constructed as in (\ref{eq:new sentence nonhomogeneous case}), corresponding to all units in all $\mathcal{A}_{n,0,j}\neq0$. 
We also introduce a set $T_{n}^{aux2}$ of all sentences $s(l,m,j;s_{k_{0}},\ldots,s_{k_{m-1}};\mu)$
built as in (\ref{eq:new sentence}), however with one modification. Namely, when considering the partition $k_{0}+\ldots+k_{m-1}=n-l$ together with $m$ sentences $s_{k_{0}}, s_{k_{1}},\ldots, s_{k_{m-1}}$, we demand that at least one of them is $s_{k_{i}}=s(k_{i},j,\alpha)$, while the rest are elements of $T_{k_{0}}^{hom}, T_{k_{1}}^{hom},\ldots, T_{k_{m-1}}^{hom}$ (all $T_{k_{i}}^{hom}$ excluded) respectively. 
Finally we define
\be
T_{n}^{nonh}=T_{n}^{hom}\cup T_{n}^{aux1}\cup T_{n}^{aux2}.
\ee
Note that sentences $s(l,j,\alpha)$ corresponding to nonhomogeneous terms are present in the recursion for $T_{n}^{nonh}$ in two ways: as themselves and as subsentences $s_{k_{i}}$, but they never contribute to their own recursion with the expression for a new sentence starting with $\textrm{sgn}(\mathcal{A}_{l,0,j})\left[\varepsilon\right]{}^{*\left(l-1\right)}*[\alpha^{*j}]$. 
Having constructed all sets $T_{n}^{nonh}$ we form
\be
T^{nonh} = \overset{\infty}{\underset{n=0}{\bigcup}}T_{n}^{nonh}.
\ee
We also consider a free algebra $B^{nonh}$, generated by elements of $T^{nonh}$.

Following section \ref{sub:Combinatorial-construction-BPS} we define now a set of primary sentences
\begin{equation}
T^{nonh,0}=\left\{ s\in T^{nonh}:\ s\neq s_{1}*s_{2}*\ldots*s_{t}\ \forall s_{i}\in T^{nonh},\ t>1\right\} \label{eq:nonhprimary}
\end{equation}
that generate the free algebra $B^{nonh,0}$. We also introduce a formal language $\left(T^{nonh,0}\right)^{*}$ over an alphabet $T^{nonh,0}$, a set of Lyndon words (which are strictly smaller than all their cyclic shifts) $T^{nonh,L}$ in this language, and a modified set
\begin{equation}
T^{nonh,L,+}=T^{nonh,L}\cup\left\{ s*s:s\in T^{nonh,L},\textrm{sgn}\left(s\right)=-1\right\} .\label{eq:T^L,+-1}
\end{equation}

There is however a crucial difference with section \ref{sub:Combinatorial-construction-BPS}, namely $B^{nonh}\ncong T\left(B^{nonh,0}\right)$. This can be seen e.g. by noting that if
\begin{equation}
s=s(l,j,\alpha)=\textrm{sgn}(\mathcal{A}_{l,0,j})\left[\varepsilon\right]{}^{*\left(l-1\right)}*[\alpha^{*j}]\in T_{l}^{nonh,0}
\end{equation}
then the tensor product of two such elements $s$ belongs to $T\left(B^{nonh,0}\right)$, but 
\begin{equation}
s*s=\left[\varepsilon\right]{}^{*\left(l-1\right)}*[\alpha^{*j}]*\left[\varepsilon\right]{}^{*\left(l-1\right)}*[\alpha^{*j}]\notin T_{2l}^{nonh}.\label{eq:s*s}
\end{equation}
This is a consequence of the fact that nonhomogeneous terms $\mathcal{A}_{l,0}(q)x^{l}$ do not correspond to the recursion, while $\mathcal{A}_{l,m}(q)x^{l}Y(x,q)^{(m;q^{2})}$ do. In other words, there are ``too many words''
in $\left(T^{nonh,0}\right)^{*}$ and this set cannot be identified with $T^{nonh}$. We can still define a map $\varphi$ that translates words from $\left(T^{nonh,0}\right)^{*}$ into sentences from $\Sigma^{**}$, analogously to (\ref{eq:phidef}) and (\ref{eq:phiconc}), however $T^{nonh}$ is a subset of the image of $\varphi$. To fix this we introduce an equivalence relation on words in $\left(T^{nonh,0}\right)^{*}$, by imposing that two words $w_{1},w_{2}\in\left(T^{nonh,0}\right)^{*}$ are equivalent, $w_{1}\sim w_{2}$, if their factorizations differ by Lyndon words whose images under $\varphi$:
\begin{itemize}
\item are not primary sentences and the first subsentence is $s(l,j,\alpha)$
from (\ref{eq:new sentence nonhomogeneous case}) for some $l,j,\alpha$, or
\item are the second or next copies of $s(l,j,\alpha)$ from (\ref{eq:new sentence nonhomogeneous case})
for some $l,j,\alpha$ in the image of factorization.
\end{itemize}
For example $w_{1}=s$ from (\ref{eq:new sentence nonhomogeneous case}) is in relation with $w_{2}=s*s$ from (\ref{eq:s*s}), because their factorizations differ by $s$, whose image under $\varphi$ is the sentence $s(l,j,\alpha)$ that appears the second time in the image of factorization. We can interpret this equivalence relation as trivializing these words in $T^{nonh,L}$, whose images under $\varphi$ would have arised in the recursion corresponding to $s(l,j,\alpha)$ from (\ref{eq:new sentence nonhomogeneous case}) for some $l,j,\alpha$. 

Now we can define a bijection $\tilde{\varphi}$ that maps the conjugacy class in $\nicefrac{\left(T^{nonh,0}\right)^{*}}{\sim}$ to the image of its shortest representative under $\varphi$. $\tilde{\varphi}$  preserves the concatenation, so we can write
\begin{equation}
\nicefrac{\left(T^{nonh,0}\right)^{*}}{\sim}\cong T^{nonh},\qquad   \nicefrac{T\left(B^{nonh,0}\right)}{\sim}\cong B^{nonh}.   \label{eq:division by relation}
\end{equation}
We also define
\begin{equation}
T^{L}=\tilde{\varphi}\left(\nicefrac{T^{nonh,L}}{\sim}\right),\qquad   T^{L,+}= \tilde{\varphi}\left(\nicefrac{T^{nonh,L,+}}{\sim}\right), \qquad  T^{L,+}_{r}= T^{L,+}\cap T_{r}^{nonh}. \label{eq:T^L nonhomogeneous case}
\end{equation}
By Chen-Fox-Lyndon theorem, (\ref{eq:division by relation}), and (\ref{eq:T^L nonhomogeneous case}) imply that every word in $T^{nonh}$ can be written in a unique way as a concatenation of elements of $T^{L}$. Finally, as in (\ref{eq:Nr(q)}), we find
\be
N_{r}(q)=\sum_j N_{r,j}q^{j+1}=\frac{Q_{r}(q)}{[r]_{q^{2}}}, \qquad Q_{r}(q)=\sum_p Q_{r,p}q^{p}=\underset{s\in T_{r}^{L,+}}{\sum}\textrm{sgn}\left(s\right)q^{\textrm{wt}(s)},\label{eq:Qr final}
\ee
which determine $Y(x,q)$ as in 
(\ref{eq:Y-product two sides}), as a solution to a nonhomogeneous equation (\ref{eq:nonhom-Y})
\begin{equation}
Y(x,q) = \prod_{r=1}^{\infty} \prod_j \prod _{l=0}^{r-1} \Big(1-x^{r}q^{j+2l+1}\Big)^{-N_{r,j}}=
\prod_{r=1}^{\infty}\prod_p \left(1-q^{p}x^{r}\right)^{-Q_{r,p}}.
\end{equation}



\subsection{$Y_{n}(q)$ vs. $Q_{r}(q)$ and explicit recursions for LMOV invariants}

So far we have provided a recursive construction of a combinatorial model that yields $Y_n(q)$ and $Q_r(p)=\sum_{p=p_{min}}^{p_{max}} Q_{r,p} q^p$ on the level of generating series, so that
\begin{equation}
Y(x,q)=\overset{\infty}{\underset{n=0}{\sum}}Y_{n}(q)x^{n}=\underset{r\geq1}{\prod}\underset{p}{\prod}\left(1-q^{p}x^{r}\right)^{-Q_{r,p}}.    \label{YQ-relation}
\end{equation}
Let us point out that, apart from the recursive construction, also a direct relation between $Y_n(q)$ and $Q_r(p)$ can be given. This relation takes form
\begin{equation}
Y_{n}(q)=\underset{1v_{1}+\ldots+nv_{n}=n}{\sum}\Big(\prod_{r=1}^n \frac{Q_{r}(q)^{(v_{r})}}{v_{r}!}\Big),\label{eq:Ydependence_onQ}
\end{equation}
where
\be
Q_{r}(q)^{(v_{r})}=\sum_{u_{p_{min}}+u_{p_{min+1}}+\ldots+u_{p_{max}}=v_{r}} \left(\begin{array}{c}v_{r}\\u_{p_{min}},u_{p_{min+1}},\ldots,u_{p_{max}}\end{array}\right)\underset{j=p_{min}}{\stackrel{p_{max}}{\prod}}Q_{r,p}^{(u_{j})}q^{pu_{j}},     \label{eq:Pochhammerforpoly} 
\ee
\be
Q_{r,p}^{(u_{j})}=Q_{r,p}(Q_{r,p}+1)(Q_{r,p}+2)\ldots(Q_{r,p}+u_{j}-1). \label{eq:Pochhammer}
\ee
The relation (\ref{eq:Ydependence_onQ}) follows from a direct expansion of both sides of (\ref{YQ-relation}), as we show in appendix \ref{sec:ConnectionY-QandRecursionN}. Moreover it can be interpreted as a manifestation of the combinatorial construction presented earlier. First note, that while a rising Pochhammer symbol (\ref{eq:Pochhammer}) is a generalization of an ordinary power, the expression (\ref{eq:Pochhammerforpoly}) can be regarded as an analogous generalization of an ordinary multinomial formula 
\be
Q_{r}(q)^{v_{r}}=\sum_{u_{p_{min}}+u_{p_{min+1}}+\ldots+u_{p_{max}}=v_{r}}\left(\begin{array}{c}v_{r}\\u_{p_{min}},u_{p_{min+1}},\ldots,u_{p_{max}}\end{array}\right)\underset{j=p_{min}}{\stackrel{p_{max}}{\prod}}Q_{r,p}^{u_{j}}q^{pu_{j}}.
\ee
Similarly, the equation (\ref{eq:Ydependence_onQ}) can be interpreted as a generalization of a multinomial formula, which corresponds to elements of growing size ($1v_{1}+\ldots+nv_{n}=n$ versus ordinary $v_{1}+\ldots+v_{n}=n$) that can be multiplied only in one particular order. This is in fact the case of the Chen-Fox-Lyndon theorem, where every sentence (from $T=\varphi ((T^{0})^{*})$ in our construction) can be written in a unique way as a concatenation of elements (from $T^{L}$) weakly decreasing lexicographically. In other words, there is a one-to-one correspondence between sentences from $T_{n}$ and Lyndon factorizations, i.e. weakly decreasing concatenations of $v_{1}$ elements from $T_{1}^{L}$, $v_{2}$ elements from $T_{2}^{L}$, etc.,
such that $1v_{1}+\ldots+nv_{n}=n$.
Equation (\ref{eq:Ydependence_onQ}) is simply a signed and weighted sum over two sides of this correspondence: a summation over elements of $T^{L}$ gives $\sum_{s\in T_{n}}\textrm{sgn}(s)q^{\textrm{wt}(s)}=Y_n(q)$, while a signed and weighted sum over all Lyndon factorizations gives the right hand side of (\ref{eq:Ydependence_onQ}).


Furthermore, the result (\ref{eq:Ydependence_onQ}) can be also transformed into an explicit recursion relation for LMOV invariants $N_{r,j}$. We present an example of such a recursion relation in section \ref{sub:Explicit recursion}.


\subsection{Classical limit}

In the classical limit  $q\to 1$ the equation (\ref{eq:A(x,Y(x,q),q)}) reduces to  an algebraic curve
\begin{equation}
\sum_{l,m}\mathcal{A}_{l,m}(1)x^{l}Y(x,1)^{m}=\mathcal{A}(x,Y(x)) = 0,
\end{equation}
where $\mathcal{A}(x,Y(x))$ is an extremal dual A-polynomial \cite{Garoufalidis:2015ewa}. A solution of this equation $Y(x)=Y(x,1)=\prod_r (1-x^{r})^{-rb_{r}}$  encodes classical LMOV invariants (\ref{y-br}), which we can now interpret as a net number of elements in $T_{r}^{L,+}$ divided by $r$
\begin{equation}
b_{r}=\lim_{q\to 1}N_{r}(q)=\frac{1}{r}\underset{s\in T_{r}^{L,+}}{\sum}\textrm{sgn}\left(s\right).\label{eq:b_r}
\end{equation}


\section{Examples}   \label{sec-examples}

In this section we present several examples of combinatorial models associated to knots. We take advantage of formulas for normalized (reduced) colored superpolynomials $P_r^{K,norm}(a,q,t)$ derived in \cite{superA,FGSS,Nawata}. To determine LMOV invariants we need to consider unreduced polynomials
\be
P^K_r(a,q)=P^{\bf 0_1}_r(a,q)P^{K,norm}_r(a,q,-1),    \label{P-normalize}
\ee
where colored HOMFLY polynomials for the unknot take form 
\be
P^{{\bf 0_1}}_{r}(a,q) = a^{-r}q^{r} \frac{(a^2,q^2)_{r}}{(q^2,q^2)_{r}}   \label{Punknot} 
\ee
where $(x,q)_r=\prod_{k=0}^{r-1}(1-x q^k)$. $P^{{\bf 0_1}}_{r}(a,q)$ satisfy the recursion relation
 \begin{equation}
P_{r+1}^{\mathbf{0_{1}}}(a,q)=\frac{aq^{r-1}-a^{-1}q^{1-r}}{q^{r}-q^{-r}}P_{r}^{\mathbf{0_{1}}}(a,q).
\end{equation}
Following section \ref{sec-review} this relation yields the equation 
\begin{equation}
\left(1-\hat{y}^{2}-\hat{x}\left(a^{-1}q-aq\hat{y}^{2}\right)\right)P^{{\bf 0_{1}}}(x,a,q)=0   \label{eq:AP=00003D0 for unknot}
\end{equation}
for the generating series of colored HOMFLY polynomials, which can be written as
\be
P^{\bf 0_1}(x,a,q)=\sum_r P^{{\bf 0_1}}_{r}(a,q) x^r = \prod_{l=0}^{\infty} \Big(1-x^{1}a^{-1}q^{0+2l+1}\Big)^{-1}\Big(1-x^{1}a^{1}q^{0+2l+1}\Big)^{1},
\ee
and which encodes just two LMOV invariants $N_{1,-1,0}=-1$ and $N_{1,1,0}=1$. When discussing extremal invariants, colored polynomials should be normalized by extremal colored polynomials of the unknot, which can be read off from (\ref{Punknot})
\be
P_{r}^{{\bf 0_{1}^{-}}}(q)=\frac{q^{r}}{(q^{2};q^{2})_{r}},\qquad
P_{r}^{{\bf 0_{1}^{+}}}(q)=\frac{(-1)^{r}q^{r^{2}}}{(q^{2};q^{2})_{r}}.    \label{unknot-extremal}
\ee


\subsection{Twist knots and $q$-Fuss-Catalan numbers}   \label{sub:Knots-with-single}

There is a large class of knots, whose extremal normalized colored polynomials take form
\begin{equation}
P_{r}^{norm,m_{\pm}}(q)=q^{r(r-1)m_{\pm}}\left(-1\right){}^{rm_{\pm}}.  \label{Pnorm-m}
\end{equation}
In the language of \cite{Gukov:2011ry} these are knots whose homological diagrams have a single generator in a top or bottom row (i.e. corresponding to a maximal or minimal degree of variable $a$). Including the unknot normalization (\ref{unknot-extremal}) and introducing $m=m_{-}$ for the minimal case and $m=m_{+}+1$ for the maximal one, unnormalized polynomials corresponding to (\ref{Pnorm-m}) read
\begin{equation}
P_{r}^{m}(q)=(-1)^{rm}\frac{q^{r^{2}m-r(m-1)}}{(q^{2};q^{2})_{r}}.   \label{Punnorm-m}
\end{equation}
In fact many objects are characterized by this expression. First, twist knots $K_p$ form one class of knots whose extremal colored polynomials take form (\ref{Punnorm-m}). In this case $p=-1,-2,-3,\ldots$ denotes $4_1,6_1,8_1,\ldots$ knots and their maximal invariants correspond to $m=m_{+}+1=2|p|+1$, while $p=1,2,3,\ldots$ denotes $3_{1},5_{2},7_{2},\ldots$ knots whose maximal invariants  correspond to $m=m_{+}+1=2p+2$. Minimal invariants for all twist knots $K_p$ with $p<0$ correspond to $m=m_-=-2$, however minimal invariants for twist knots with $p>0$ are not of the form (\ref{Punnorm-m}). In addition $m=0,1$ correspond respectively to the minimal and maximal colored polynomials for the unknot. The case $m=2$ does not correspond to any knot, however it is related to a certain (non-standard) $q$-deformation of Catalan numbers, and similarly arbitrary $m$ is related to a (non-standard) $q$-deformation of Fuss-Catalan numbers. Values of $m$ corresponding to twist knots are summarized in table \ref{table:k for different knots}. Also colored extremal polynomials of the framed unknot have form (\ref{Punnorm-m}). Finally, analogous formulas appear in the context of Donaldson-Thomas invariants for $m$-loop quivers in \cite{Rei12}, and we will take advantage of these results to show integrality of LMOV invariants corresponding to (\ref{Punnorm-m}). 

\begin{table}[h]
\begin{centering}
\caption{Values of $m$ for maximal and minimal (denoted $\pm$ respectively) invariants for various knots.}
\label{table:k for different knots}
\par\end{centering}

\centering{}%
\begin{tabular}{|c||c|c|c|c|c|c|c|c|c|c|c|c|}
\hline 
$m$ & $-2$ & 0 & 1 & 2 & 3 & 4 & 5 & 6 & 7 & 8 & 9 & 10\tabularnewline
\hline 
case & $K^-_p, p<0$ & $0_{1}^{-}$ & $0_{1}^{+}$ & Catalan & $4_{1}^{+}$ & $3_{1}^{+}$ & $6_{1}^{+}$ & $5_{2}^{+}$ & $8_{1}^{+}$ & $7_{2}^{+}$ & $10_{1}^{+}$ & $9_{2}^{+}$\tabularnewline
\hline 
\end{tabular}
\end{table}

Note that $P_{r}^{m}(q)$ satisfies a recursion relation of the form
\begin{equation}
\big(1-q^{2(r+1)}\big)P_{r+1}^{m}(q)=(-1)^{m}q^{2rm+1}P_{r}^{m}(q),
\end{equation}
which equivalently can be written as
\begin{equation}
\widehat{A}^{m}(\widehat{M},\widehat{L},q)P_{r}^{m}(q)=0, \qquad 
\widehat{A}^{m}(\widehat{M},\widehat{L},q)=\widehat{L}-\widehat{M}^{2}\widehat{L}-q (-\widehat{M}^{2} )^{m}.
\end{equation}
Upon redefinitions discussed in section \ref{sec-review} we find a difference equation 
\begin{equation}
\widehat{\mathcal{A}}^{m}(\widehat{x},\widehat{y},q)P^{m}(x,q)=0, \qquad 
\widehat{\mathcal{A}}^{m}(\widehat{x},\hat{y},q)=1-\widehat{y}^{2}-q\widehat{x}\left(-\widehat{y}^{2}\right)^{m}   \label{calAxyP-1}
\end{equation}
for the generating series $P^{m}(x,q)=\sum_{r=0}^{\infty}P_{r}^{m}(q)x^{r}$. In terms of $Y^{m}(x,q)=\frac{P^{m}(q^{2}x,q)}{P^{m}(x,q)}$ defined in (\ref{eq:Y def}) this can be rewritten in the form (\ref{eq:A(x,Y(x,q),q)}) 
\begin{equation}
\mathcal{A}^{m}(x,Y^{m}(x,q),q)=1-Y^{m}(x,q)-qx\left(-Y^{m}(x,q)\right)^{(m;q^{2})}=0.  \label{AxYq-m} 
\end{equation}
This equation is of the homogeneous form (\ref{eq:Equation for Y in good class for recursion}). Let us focus on non-negative $m$; in this case the corresponding recursion relation (\ref{eq:Recursion}) reads
\begin{equation}
Y_{n}^{m}(q)=\left(-1\right)^{m+1}q\sum_{k_{0}+\ldots+k_{m-1}=n-1}\Big(\underset{i=0}{\overset{m-1}{\prod}}q^{2ik_{i}}Y_{k_{i}}^{m}(q)\Big)    \label{eq:Recursion for m}
\end{equation}
and we can construct a combinatorial model, following the prescription presented in sections \ref{sub:Combinatorial-construction for coefficients of Y} and \ref{sub:Combinatorial-construction-BPS}. Since in this case $\sum_{l\geq1,m\geq1,j}\left|\mathcal{A}_{l,m,j}\right|=1$, the alphabet $\Sigma$ consists of one letter and all words and sentences in the model are built out of a unique one letter word $\mu$.

To construct a set $T$, we start with the initial data $T_{0}=\{[\,]\}$ and recursively build sets $T_{n}$. Having fixed a partition $k_{0}+\ldots+k_{m-1}=n-1$ and $m$ sentences $s_{k_{0}}, s_{k_{1}},\ldots, s_{k_{m-1}}$ from $T_{k_{0}}, T_{k_{1}},\ldots,T_{k_{m-1}}$ respectively,  we define a new sentence according to (\ref{eq:new sentence})

\begin{equation}
s(1,m,1;s_{k_{0}},\ldots,s_{k_{m-1}};\mu)=\label{eq:new sentence - m}
\end{equation}
\[
=\left(-1\right)^{m+1}\left[\mu\right]*\left([\mu^{*2\left(m-1\right)}]{}^{*k_{m-1}}\vee s_{k_{m-1}}\right)*\cdots*\left([\mu^{*2}]{}^{*k_{1}}\vee s_{k_{1}}\right)*s_{k_{0}}.
\]
$T_{n}$ is a set of all $s(1,m,1;s_{k_{0}},\ldots,s_{k_{m-1}};\mu)$ for all choices of partitions $k_{0}+\ldots+k_{m-1}=n-1$ and sentences $s_{k_{0}}, s_{k_{1}}, \ldots, s_{k_{m-1}}$ from $T_{k_{0}}, T_{k_{1}},\ldots ,T_{k_{m-1}}$ respectively. Note that the sign of an $n$-word sentence is $\left(-1\right)^{n(m+1)}$.

In this example there is only one letter and the whole information is encoded in powers of $\mu$, which correspond to numbers of letters in words. Therefore we can simplify the notation and identify a sentence $s=\left(-1\right)^{n(m+1)}[\mu^{*\phi_{1}},\mu^{*\phi_{2}},\ldots,\mu^{*\phi_{n}}]$ with a signed list
\begin{equation}
\phi=\left(-1\right)^{n(m+1)}\left[\phi_{1},\phi_{2},\ldots,\phi_{n}\right]
\end{equation}
of weight $\textrm{wt}(\phi)=\textrm{wt}(s)=\phi_{1}+\phi_{2}+\cdots+\phi_{n}$. Now (\ref{eq:new sentence - m}) is equivalent to 
\begin{equation}
\phi(m;\phi_{\left(k_{0}\right)},\ldots,\phi_{\left(k_{m-1}\right)})=\label{eq:number representation}
\end{equation}
\[
=\left(-1\right)^{m+1}\left[1\right]*\Big([\underset{k_{m-1}}{\underbrace{2\left(m-1\right),\ldots,2\left(m-1\right)}}]\vee\phi_{\left(k_{m-1}\right)}\Big)*\cdots*\Big([\underset{k_{1}}{\underbrace{2,\ldots,2}}]\vee\phi_{\left(k_{1}\right)}\Big)*\phi_{\left(k_{0}\right)},
\]
with all operations on lists inherited from respective operations on sentences (note that $\phi_{\left(k_{i}\right)}$ denotes a list equivalent to the sentence $s_{k_{i}}$ from $T_{k_{i}}$, whereas $\phi_{i}$ denotes the $i$-th element from the list $\phi$). Using (\ref{eq:number representation}) recursively we can construct a set of all $n$-element lists $\phi=(-1)^{(m+1)n}\left[\phi_{1},\phi_{2},\ldots,\phi_{n}\right]$
such that
\be
\phi_{1}=1,\qquad  \phi_{i}\geq1\ \forall i=1,2,...,n, \qquad  \phi_{i+1}-\phi_{i} \in 2\mathbb{Z},\qquad \phi_{i+1}-\phi_{i}\leq2(m-1).
\ee
If $\phi$ satisfies these conditions we call it a \emph{maximally-$2(m-1)$-step list}. It follows that $T_{n}$ is equivalent to the set of all maximally-$2(m-1)$-step lists of $n$ elements, and $T=\bigcup_n T_n$ is the set of all maximally-$2(m-1)$-step lists.

Furthermore, following (\ref{eq:T zero definition}), we call a list $\phi$ is \emph{primary} if
\begin{equation}
\phi\neq\phi_{(1)}*\phi_{(2)}*\ldots*\phi_{(t)}\quad \forall\phi_{(i)}\in T,\ t>1,
\end{equation}
and $T^{0}=\bigcup_n  T_{n}^{0}$ is a set of all primary maximally-$2(m-1)$-step lists. We can build $T^0$ recursively. Note that a list $\phi$ is primary if and only if it is of the form
\begin{equation}
\phi=\phi(m,[],\phi_{\left(k_{1}\right)},\ldots,\phi_{\left(k_{m-1}\right)})=\label{eq:recursion primary}
\end{equation}
\[
=\left(-1\right)^{m+1}\left[1\right]*\Big([\underset{k_{m-1}}{\underbrace{2\left(m-1\right),\ldots,2\left(m-1\right)}}]\vee\phi_{\left(k_{m-1}\right)}\Big)*\cdots*\Big([\underset{k_{1}}{\underbrace{2,\ldots,2}}]\vee\phi_{\left(k_{1}\right)}\Big),
\]
so it corresponds to the partition $k_{0}=0,\ k_{1}+k_{2}+\ldots+k_{m-1}=n-1$. In consequence we can write that $T_{n}^{0}$ is a set of all $\phi(m,[],\phi_{\left(k_{1}\right)},\ldots,\phi_{\left(k_{m-1}\right)})$ for all choices of partitions $k_{1}+\ldots+k_{m-1}=n-1$ and lists $\phi_{\left(k_{1}\right)},\dots,\phi_{\left(k_{m-1}\right)}$ from $T_{k_{1}},\dots ,T_{k_{m-1}}$ respectively. Having constructed $T^{0}$, we identify it as a new alphabet with the ordering induced from $\mathbb{N}$. Following section \ref{sub:Combinatorial-construction-BPS} we define also the set of Lyndon words $T^{L},$ and $T^{L,+}$, and ultimately LMOV invariants are given as in (\ref{eq:Nr(q)})
\begin{equation}
N_{r}^{m}(q)=\underset{j}{\sum}N_{r,j}^{m}q^{j+1}=\frac{1}{[r]_{q^{2}}}\underset{\phi\in T_{r}^{L,+}}{\sum}(-1)^{(m+1)r}q^{\textrm{wt}(\phi)}.\label{eq:Nr(q)-1-1}
\end{equation}

Interestingly, the model discussed above, associated to colored polynomials (\ref{Punnorm-m}), is equivalent to the combinatorics of the degenerate Cohomological Hall algebra of the $m$-loop quiver considered in \cite{Rei12}. In particular it is proved in \cite{Rei12} that Donaldson-Thomas invariants for $m$-loop quiver $\textrm{DT}_{r}^{(m)}(q)$ determined from such a model are integer, and in appendix \ref{sec:Duality us-Reineke} we show that these invariants are related to our LMOV invariants by a simple redefinition $N_{r}^{m}(q)=(-1)^{(m+1)r}q^{3r-2}\textrm{DT}_{r}^{(m)}(q^{2})$. In consequence LMOV invariants $N_{r,j}^{m}$ are integer too, which proves the LMOV conjecture in the extremal case for a large class of knots. Moreover, in the classical limit this proves divisibility statements presented in \cite{Garoufalidis:2015ewa}.


\subsection{Explicit recursion relation for LMOV invariants}\label{sub:Explicit recursion}

From the knowledge of the dual A-polynomial equation one can determine an explicit recursion relation for LMOV invariants. We illustrate this statement in the case of the equation of the form (\ref{AxYq-m}), corresponding to colored polynomials in (\ref{Punnorm-m}). Let us consider first a non-negative integer $m$ and consider combinations $N_{r}^{m}(q)=\sum_j N^m_{r,j}q^{j+1}$ introduced in (\ref{Nrq-def}). In this case
\begin{equation}
Y^{m}(x,q)^{(m;q^{2})}=\prod_{r=1}^{\infty} \prod_{l=0}^{mr-1} \prod_{j=j_{min}}^{j_{max}}\Big(1-x^{r}q^{j+1+2l}\Big)^{-N_{r,j}^{m}},
\end{equation}
so that (\ref{eq:Ydependence_onQ}) yields 
\begin{equation}
Y^{m}(x,q)^{(m;q^{2})}=\sum_{n=0}^{\infty} \bigg( \sum_{1v_{1}+2v_{2}+\ldots+nv_{n}=n}\Big(\underset{r=1}{\stackrel{n}{\prod}}\frac{1}{v_{r}!}\left(N_{r}^{m}(q)[mr]_{q^{2}}\right)^{(v_{r})}\Big) \bigg) x^{n}.
\end{equation}
The leading term (of order $x^0$) in (\ref{AxYq-m}) reads $1-Y^{m}(0,q)=0$, which implies the initial condition $N_{0}^{m}(q)=1$. Comparing coefficients at higher powers of $x$ we get a relation
\begin{equation}
\begin{split}
\label{eq:comparison of coeffs}
\underset{1v_{1}+\ldots+nv_{n}+(n+1)v_{n+1}=n+1}{\sum}&\Big(\underset{r=1}{\stackrel{n+1}{\prod}}\frac{1}{v_{r}!}\left(N_{r}^{m}(q)[r]_{q^{2}}\right)^{(v_{r})}\Big) = \\
& =(-1)^{m+1}q\underset{1v_{1}+\ldots+nv_{n}=n}{\sum}\Big(\underset{r=1}{\stackrel{n}{\prod}}\frac{1}{v_{r}!}\left(N_{r}^{m}(q)[mr]_{q^{2}}\right)^{(v_{r})}\Big).
\end{split}
\end{equation}
The condition $1v_{1}+\ldots+nv_{n}+(n+1)v_{n+1}=n+1$ in the summation in the left hand side is satisfied either for $v_{n+1}=1$ and
$v_1=\ldots=v_n=0$, or for $v_{n+1}=0$ and $1v_{1}+\ldots+nv_{n}=n+1$. It follows that  (\ref{eq:comparison of coeffs}) can be written as
\begin{equation}
\begin{split}
N_{n+1}^{m}(q)=-\frac{1}{[n+1]_{q^{2}}}\bigg(&\sum_{1v_{1}+\ldots+nv_{n}=n+1}\Big(\prod_{r=1}^n \frac{1}{v_{r}!} (N_{r}^{m}(q)[r]_{q^{2}} )^{(v_{r})}\Big)+\\
&+(-1)^{m}q\sum_{1v_{1}+\ldots+nv_{n}=n}\Big(\prod_{r=1}^n \frac{1}{v_{r}!} (N_{r}^{m}(q)[mr]_{q^{2}} )^{(v_{r})}\Big)\bigg),   \label{eq:explicit recursion}
\end{split}
\end{equation}
which constitutes an explicit recursion relation for $N_{r}^{m}(q)$.


Analogous computation for the equation (\ref{AxYq-m}) with negative $m$, for 
\be
Y(x,q)^{(m;q^{2})}=\frac{\widehat{y}^{-2|m|}P(x,q)}{P(x,q)}=\frac{P(q^{-2|m|}x,q)}{P(x,q)}=\underset{i=1}{\overset{|m|}{\prod}}Y(q^{-2i}x,q)^{-1},
\ee
leads to the same initial condition $N_{0}^{m}(q)=1$ and a recursion of the form
\begin{equation}
\begin{split}
\label{eq:explicit recursion negative m}
N_{n+1}^{m}(q)=-\frac{1}{[n+1]_{q^{2}}}\bigg(&\sum_{1v_{1}+\ldots+nv_{n}=n+1}\Big(\prod_{r=1}^n  \frac{1}{v_{r}!}\left(N_{r}^{m}(q)[r]_{q^{2}}\right)^{(v_{r})}\Big)+\\
&+(-1)^{m}q\sum_{1v_{1}+\ldots+nv_{n}=n}\Big(\prod_{r=1}^n  \frac{1}{v_{r}!}\left(-N_{r}^{m}(q)q^{2mr}[|m|r]_{q^{2}}\right)^{(v_{r})}\Big)\bigg). 
\end{split}
\end{equation}
For example the minimal dual A-polynomial for twist knots $K_p$ with $p<0$ knot reads $\widehat{\mathcal{A}}(\widehat{x},\widehat{y},q)=\widehat{y}^{4}-\widehat{y}^{6}-q^{5}\widehat{x}$, or equivalently it takes form (\ref{calAxyP-1}) with $m=-2$, which imposes an equation $(1-\widehat{y}^{2}+q\widehat{x}\widehat{y}^{-4})P(x,q)=0$, so that (\ref{AxYq-m}) takes form
\begin{equation}
1-Y(x,q)-qxY(x,q)^{(-2;q^{2})}=0.
\end{equation}
Solving then (\ref{eq:explicit recursion negative m}) with $m=-2$ we find that $N_{r}^{-2}(q)$ have integer coefficients and encode correct extremal LMOV invariants for $K_p$ knots with $p<0$, for example
\be 
N_{1}^{-2}(q)=-\frac{q}{[1]_{q^{2}}}=-q,\qquad
N_{2}^{-2}(q)=-\frac{q^{-2}+1}{[2]_{q^{2}}}=-q^{-2},\qquad 
N_{3}^{-2}(q) 
=-q^{-3}-q^{-5}-q^{-9}.  
\ee


\subsection{$m=2$ and novel $q$-deformed Catalan numbers} \label{sub:Catalan-case}

As a more specific example let us consider (\ref{Punnorm-m}) with $m=2$. This case does not correspond to any twist knot, however it provides a certain non-standard $q$-deformation of Catalan numbers, which is interesting in its own right. In this case the equation (\ref{AxYq-m}) takes form
\begin{equation}
\mathcal{A}(x,Y(x,q),q)=1-Y(x,q)-qx\left(-Y(x,q)\right)^{(2;q^{2})}=0,    \label{eq:A_k(x,Y)-1}
\end{equation}
and it leads to the following recursion relation for $C_n(q)=(-1)^n Y_n(q)$ 
\begin{equation}
C_{n}(q)=\sum_{k_{0}+k_{1}=n-1} q^{2k_{1}+1} C_{k_{0}}(q)C_{k_{1}}(q)      \label{Catalan-new}
\end{equation}
with the initial condition $C_0(q)=1$. In the classical limit (\ref{eq:A_k(x,Y)-1}) reduces to $1-Y(x)-xY(x)^{2}=0$, which is also the classical dual A-polynomial equation for $5_2$ knot (however the quantum equation (\ref{eq:A_k(x,Y)-1}) does not encode quantum LMOV invariants for this knot). For $Y(x)=\sum_n Y_n x^n$, the coefficients $C_n=(-1)^n Y_n$ are ordinary Catalan numbers that satisfy $C_n=\sum_{k_0+k_1=n-1} C_{k_0}C_{k_1}$, hence $C_n(q)$ can be regarded as $q$-deformed Catalan numbers. The crucial property of $C_n(q)$ is that they encode integer invariants $N_{r,j}$ through (\ref{eq:Y def}), which is not the case for another, more standard $q$-deformation of Catalan numbers $c_n(q)$ defined via $c_{n}(q)=\sum_{k_{0}+k_{1}=n-1}  q^{k_{1}} c_{k_{0}}(q) c_{k_{1}}(q)$ \cite{FURLINGER1985248}, so that $C_n(q)=q^n c_n(q^2)$.

\begin{figure}[h]
\begin{center}
\includegraphics[width=25mm]{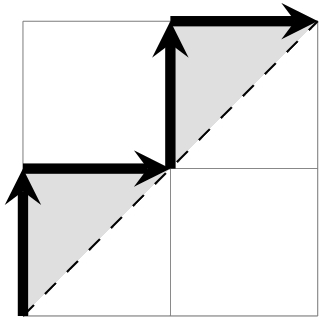}$\qquad \qquad$
\includegraphics[width=25mm]{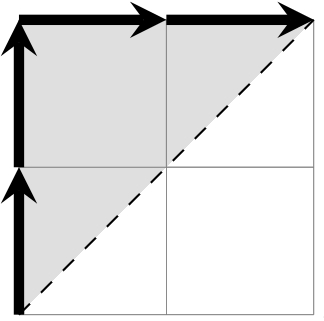}
\end{center}
\caption{Dyck paths representation of $q$-deformed Catalan numbers.}
\label{figCatalans}
\end{figure}

Let us construct now a combinatorial model, following the prescription given in section \ref{sub:Knots-with-single}. As $m=2$, $T_{n}$ is a set of all maximally-$2$-step lists of $n$ elements and sign $(-1)^{n}$, which can be represented in terms of lists or column of boxes (all filled with $\mu$, which we suppress):
\begin{align}
T_{1}&=\{-[1]\}=\left\{ -\young(\hfil)\right\} \nonumber \\
T_{2}&=\{[1,1],[1,3]\}=\left\{ \young(\hfil\hfil),\young(:\hfil,:\hfil,\hfil\hfil)\right\}  \nonumber \\
T_{3}&=\{-[1,1,1],\,-[1,1,3],\,-[1,3,1],\,-[1,3,3],\,-[1,3,5]\}=  \nonumber \\
&=\left\{ -\young(\hfil\hfil\hfil),\,-\young(::\hfil,::\hfil,\hfil\hfil\hfil),\,-\young(:\hfil:,:\hfil:,\hfil\hfil\hfil),\,-\young(:\hfil\hfil,:\hfil\hfil,\hfil\hfil\hfil),\,-\young(::\hfil,::\hfil,:\hfil\hfil,:\hfil\hfil,\hfil\hfil\hfil)\right\} \nonumber
\end{align}
The above representation can be also translated to a familiar representation of Catalan numbers in terms of Dyck paths, i.e. paths above a diagonal in a square grid, connecting bottom left and top right corners of the square. In this case columns in the above pictures correspond to rows in the grid of the Dyck path, and every box $\young(\hfil)$ is translated to one triangle left to the diagonal, respectively of the form \includegraphics[width=4mm]{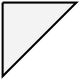} and
\includegraphics[width=4mm]{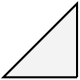}.
For example $[1,1]=\young(\hfil\hfil)$ corresponds to two triangles \includegraphics[width=4mm]{uppertriangle.jpg} in two rows, so the resulting Dyck path is shown in the left in figure \ref{figCatalans}. Similarly $[1,3]=\young(:\hfil,:\hfil,\hfil\hfil)$ corresponds to one triangle \includegraphics[width=4mm]{uppertriangle.jpg} in the bottom row and three triangles \includegraphics[width=4mm]{uppertriangle.jpg}, \includegraphics[width=4mm]{lowertriangle.jpg}, and \includegraphics[width=4mm]{uppertriangle.jpg} in the top row, and the resulting Dyck path is shown in the right in figure \ref{figCatalans}.
Ordinary Catalan numbers $C_n$ are given by the number of Dyck paths in a square of size $n$. In addition the exponent of $q$ in ordinary $q$-Catalans $c_n(q)$ counts the number of full squares \includegraphics[width=4mm]{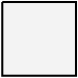} above the diagonal and restricted by a given path, whereas for $C_n(q)$ defined via (\ref{Catalan-new}) the power of $q$ counts all over-diagonal triangles \includegraphics[width=4mm]{uppertriangle.jpg} or \includegraphics[width=4mm]{lowertriangle.jpg}.

Let us illustrate other ingredients of the combinatorial construction. $T^{0}$ in the Catalan case is a set
of all primary maximally-$2$-step lists, and its subsets with up to 3 elements are
\begin{align}
T_{1}^{0}&=\{-[1]\}=\left\{ -\young(\hfil)\right\} \\
T_{2}^{0}&=\{[1,3]\}=\left\{\young(:\hfil,:\hfil,\hfil\hfil)\right\} \\
T_{3}^{0}&=\{-[1,3,3],\,-[1,3,5]\}=\left\{-\young(:\hfil\hfil,:\hfil\hfil,\hfil\hfil\hfil),\,-\young(::\hfil,::\hfil,:\hfil\hfil,:\hfil\hfil,\hfil\hfil\hfil)\right\}
\end{align}
Lyndon words in $\left(T^{0}\right)^{*}$ that are lists of length 1, 2, and 3 take form
\begin{align}
T_{1}^{L}&=\{-[1]\}=\left\{ -\young(\hfil)\right\}   \\
T_{2}^{L}&=\{[1,3]\}=\left\{\young(:\hfil,:\hfil,\hfil\hfil)\right\}    \\
T_{3}^{L}&=\{-[1,1,3],\,-[1,3,3],\,-[1,3,5]\}=\left\{ -\young(::\hfil,::\hfil,\hfil\hfil\hfil),\,-\young(:\hfil\hfil,:\hfil\hfil,\hfil\hfil\hfil),\,-\young(::\hfil,::\hfil,:\hfil\hfil,:\hfil\hfil,\hfil\hfil\hfil)\right\}
\end{align}
Note that $T$ contains lists that arise as a concatenation of a pair of the same lists of negative sign, for example $[1,1]=\left(-[1]\right)*\left(-[1]\right)$. In consequence we have
\begin{align}
T_{1}^{L,+}&=\{-[1]\}=\left\{ -\young(\hfil)\right\}  \\
T_{2}^{L,+}&=\{[1,1],[1,3]\}=\left\{ \young(\hfil\hfil),\young(:\hfil,:\hfil,\hfil\hfil)\right\}  \\
T_{3}^{L,+}&=\{-[1,1,3],\,-[1,3,3],\,-[1,3,5]\}=\left\{ -\young(::\hfil,::\hfil,\hfil\hfil\hfil),\,-\young(:\hfil\hfil,:\hfil\hfil,\hfil\hfil\hfil),\,-\young(::\hfil,::\hfil,:\hfil\hfil,:\hfil\hfil,\hfil\hfil\hfil)\right\}
\end{align} 
Counting the number of boxes in the above pictures and including signs we get
\be
N_{1}(q)=\frac{-q}{[1]_{q^{2}}}=-q,  \qquad  N_{2}(q)=\frac{q^{2}+q^{4}}{[2]_{q^{2}}}=q^{2}, \qquad  N_{3}(q)=\frac{-q^{5}-q^{7}-q^{9}}{[3]_{q^{2}}}=-q^{5}, 
\ee
thus BPS invariants $N_{r,j}$ for $r=1,2,3$ take form $N_{1,0}=-1, N_{2,1}=1, N_{3,4}=-1$.


\subsection{$m=3$ and $4_{1}$ knot}

Let us illustrate the construction from section \ref{sub:Knots-with-single} also in the case $m=3$, which corresponds to maximal invariants for $4_{1}$ knot. In this case the equation (\ref{AxYq-m}) reads
\begin{equation}
\mathcal{A}(x,Y(x,q),q)=1-Y(x,q)+qxY(x,q)^{(3;q^{2})}=0\label{eq:A^{4_1+}(x,Y)}
\end{equation}
and the recursion relation (\ref{eq:Recursion for m}) takes form
\begin{equation}
Y_{n}(q)=\sum_{k_{0}+k_{1}+k_{2}=n-1}   q^{2k_1+4k_2+1}Y_{k_{0}}(q) Y_{k_{1}}(q) Y_{k_{2}}(q).   \label{eq:Recursion for 4_1^+}
\end{equation}
As explained in section \ref{sub:Knots-with-single} in this case there is a unique one letter word $\mu$, and since $m+1=4$ is even, the sign of all sentences is positive. 

The initial condition in the construction of the set $T$ is $T_{0}=\{[\,]\}  =\left\{ \cdot\right\}$, denoting respectively an empty list and (in a graphical representation) no box. For $n=1$ there is only one partition $k_{0}=k_{1}=k_{2}=0$ of $n-1$  and $T_{k_{0}}=T_{k_{1}}=T_{k_{2}}=T_{0}$ contains only one sentence $\left[\,\right] = s_{k_{0}}=s_{k_{1}}=s_{k_{2}}$, therefore $s(1,3,1;\left[\,\right],\left[\,\right],\left[\,\right];\mu)=\left[\mu\right]*\left[\,\right]*\left[\,\right]*\left[\,\right]=[\mu]$ 
and
\begin{equation}
T_{1}=\{[\mu]\}=\left\{ \young(\mu)\right\} \label{eq:T1}
\end{equation}
or equivalently $T_1=\{[1]\}$ using the notation of lists introduced in section \ref{sub:Knots-with-single}. 

For $n=2$ there are three partitions of $n-1$: $(k_{0}=1,\, k_{1}=0,\, k_{2}=0)$, $(k_{0}=0,\, k_{1}=1,\, k_{2}=0)$, and  $(k_{0}=0,\, k_{1}=0,\, k_{2}=1)$, so new sentences take form
\begin{equation}\label{eq:new sentence - n one}
\begin{split}
s(1,3,1;\left[\mu\right],\left[\,\right],\left[\,\right];\mu)&=\left[\mu\right]*\left[\,\right]*\left[\,\right]*\left[\mu\right]=[\mu,\mu]\\
s(1,3,1;\left[\,\right],\left[\mu\right],\left[\,\right];\mu)&=\left[\mu\right]*\left[\,\right]*\left([\mu^{*2}]^{*1}\vee\left[\mu\right]\right)*\left[\,\right]=[\mu,\mu^{*3}]\\
s(1,3,1;\left[\,\right],\left[\,\right],\left[\mu\right];\mu)&=\left[\mu\right]*\left([\mu^{*4}]^{*1}\vee\left[\mu\right]\right)*\left[\,\right]*\left[\,\right]=[\mu,\mu^{*5}]\\
\end{split}
\end{equation}
and therefore $T_{2}=\{[1,1],[1,3],[1,5]\}$ in the notation of lists, or in more detail
\begin{equation}\label{eq:T2}
\begin{split}
T_{2}=\{[\mu,\mu],[\mu,\mu^{*3}],[\mu,\mu^{*5}]\} = \left\{ \young(\mu\mu),\young(:\mu,:\mu,\mu\mu),\young(:\mu,:\mu,:\mu,:\mu,\mu\mu)\right\} 
\end{split}
\end{equation}
which is the set of all maximally-$4$-step lists of $2$ elements. 

For $n=3$ there are 6 partitions of $n-1$
\be
\begin{split}
k_{0}&=2,\, k_{1}=0,\, k_{2}=0;\qquad k_{0}=0,\, k_{1}=2,\, k_{2}=0;\qquad k_{0}=0,\, k_{1}=0,\, k_{2}=2; \\
k_{0}&=1,\, k_{1}=1,\, k_{2}=0;\qquad k_{0}=1,\, k_{1}=0,\, k_{2}=1;\qquad k_{0}=0,\, k_{1}=1,\, k_{2}=1,
\end{split}
\ee
which correspond to:
\be
\begin{split}
&\phi(3;\phi_{\left(2\right)},\phi_{\left(0\right)},\phi_{\left(0\right)});\qquad\phi(3;\phi_{\left(0\right)},\phi_{\left(2\right)},\phi_{\left(0\right)});\qquad\phi(3;\phi_{\left(0\right)},\phi_{\left(0\right)},\phi_{\left(2\right)});\\
&\phi(3;\phi_{\left(1\right)},\phi_{\left(1\right)},\phi_{\left(0\right)});\qquad\phi(3;\phi_{\left(1\right)},\phi_{\left(0\right)},\phi_{\left(1\right)});\qquad\phi(3;\phi_{\left(0\right)},\phi_{\left(1\right)},\phi_{\left(1\right)}).
\end{split}
\ee
Since $T_{2}$ contains three lists, every $\phi_{(2)}$ contributes three times: once as $[1,1]$, the second time as $[1,3]$, and the last time as $[1,5]$. In consequence we have (in the notation of lists) 
\begin{equation}
\begin{split}
T_{3}=\{&[1,1,1],\,[1,1,3],\,[1,1,5],[1,3,1],\,[1,3,3],\,[1,3,5],\\
& [1,3,7],\,[1,5,1],\,[1,5,3],\,[1,5,5],\,[1,5,7],\,[1,5,9]\},
\end{split}
\end{equation}
which is the set of all maximally-$4$-step lists of $3$ elements.
Furthermore, the sets of primary maximally-$4$-step lists of $1$, $2$, and $3$ elements take form
\begin{equation}
\begin{split}
T_{1}^{0}&=\{[1]\},\qquad    T_{2}^{0}=\{[1,3],[1,5]\},\\
T^{0}_{3}&=\{[1,3,3],\,[1,3,5],\,[1,3,7],\,[1,5,3],\,[1,5,5],\,[1,5,7],\,[1,5,9]\}, 
\end{split}
\end{equation}
and then Lyndon words are
\begin{equation}
\begin{split}
T_{1}^{L}=\{&[1]\},\qquad      T_{2}^{L}=\{[1,3],[1,5]\},\\
T_{3}^{L}=\{&[1,1,3],\,[1,1,5],\,[1,3,3],\,[1,3,5],\,[1,3,7],\,[1,5,3],\,[1,5,5],\,[1,5,7],\,[1,5,9]\}.   
\end{split}
\end{equation}
Since the sign of all lists is positive, it follows from (\ref{eq:T^L,+}) that $T_{r}^{L,+}=T_{r}^{L}$, and (\ref{eq:Nr(q)}) yields
\begin{align}
N_{1}(q)&=\frac{q}{[1]_{q^{2}}}=q,\qquad   
N_{2}(q)=\frac{q^{4}+q^{6}}{[2]_{q^{2}}}=q^{4},\\
N_{3}(q)&=\frac{q^{5}+2q^{7}+2q^{9}+2q^{11}+q^{13}+q^{15}}{[3]_{q^{2}}}=q^{5}+q^{7}+q^{11}.
\end{align}
Therefore LMOV invariants take form $N_{1,0}=1$, $N_{2,3}=1$, and $N_{3,4}=N_{3,6}=N_{3,10}=1$.


\subsection{Torus knots}

In this section we find quantum dual extremal A-polynomials and determine LMOV invariants for torus knots of type $(2,2p+1)$, and present in detail a construction of a combinatorial model for the trefoil knot. For $(2,2p+1)$ torus knot colored normalized superpolynomials (labeled by appropriate $p$) take form \cite{FGSS}
\be
\begin{split}
P_r^{p,norm}(a,q,t) &= a^{2pr} q^{-2pr}  \sum_{0\le k_p \le \ldots \le k_2 \le k_1 \le r}
\left[\!\begin{array}{c} r\\k_1 \end{array}\!\right]\left[\!\begin{array}{c} k_1\\k_2 \end{array}\!\right]\cdots\left[\!\begin{array}{c} k_{p-1}\\k_p \end{array}\!\right]  \times\\
&   \times q^{2(2r+1)(k_1+k_2+\ldots+k_p)-2\sum_{i=1}^p k_{i-1}k_i} t^{2(k_1+k_2+\ldots+k_p)} \prod_{i=1}^{k_1}(1+a^2q^{2(i-2)}t),      \label{fort2k}
\end{split}
\ee
where ${n \brack k}= \frac{(q^2;q^2)_n}{(q^2;q^2)_k(q^2;q^2)_{n-k}}$. Extremal unnormalized HOMFLY polynomials $P^p_r(q)$ are obtained by including an appropriate unknot factor (\ref{unknot-extremal}), setting $t=-1$ and ignoring $a^{2pr}$; in addition in the minimal case the product $\prod_{i=1}^{k_1}(1+a^2q^{2(i-2)}t)$ must be ignored, while in the maximal case one should pick up from this product only the coefficient $(-1)^r q^{2\sum_{i=1}^r (i-2)}=(-1)^r q^{r^2-3r}$ (at the highest power of $a$) and fix $k_1=r$ in the overall expression. 

To determine dual extremal quantum A-polynomials we first find --  using \cite{qZeil} -- quantum extremal A-polynomials that annihilate (or impose recursion relations for) the above extremal colored HOMFLY polynomials, and then, as explained in section \ref{sec-review}, we determine their dual counterparts that annihilate generating series (\ref{Pminmax product form}), according to (\ref{extremal calAxyP}). Dual extremal quantum A-polynomials for $3_1$ and $5_1$ knots found in this way take form
\be
\begin{split}
\widehat{\mathcal{A}}^{+}_{3_1}(\hat{x},\hat{y},q) = & 1-\hat{y}^{2}-q\hat{x}\hat{y}^{8}, \\
\widehat{\mathcal{A}}^{-}_{3_1} (\hat{x},\hat{y},q) = & 1-q^{-1}\hat{x}-\hat{y}^{2}+q\hat{x}\hat{y}^{2}-q\hat{x}\hat{y}^{4}-q^{3}\hat{x}\hat{y}^{4}-q^{6}\hat{x}^{2}\hat{y}^{6}, \\
\widehat{\mathcal{A}}^{+}_{5_1}(\hat{x},\hat{y},q) = & 1-\hat{y}^{2}-q^{-1}\hat{x}\hat{y}^{8}+q\hat{x}\hat{y}^{10} -q(1+q^{2})\hat{x}\hat{y}^{12}-q^{14}\hat{x}^{2}\hat{y}^{22}, \\
\widehat{\mathcal{A}}^{-}_{5_1}(\hat{x},\hat{y},q) = &  1-q^{-3}\hat{x}-(1-q^{-1}\hat{x})\hat{y}^{2}-q^{-1}(1+q^{2})\hat{x}\hat{y}^{4}+q^{-1}(1+q^{2}+q^{4}-q^{3}\hat{x})\hat{x}\hat{y}^{6}+\\
&-q^{-1}(1+q^{2}+q^{4}+q^{6}-q^{5}\hat{x})\hat{x}\hat{y}^{8}-q^{4}(1+2q^{2}+q^{4})\hat{x}^{2}\hat{y}^{10}+\\
&+q^{6}(1+q^{2}+q^{4})\hat{x}^{2}\hat{y}^{12}-q^{6}(1+q^{2}+2q^{4}+q^{6}+q^{8})\hat{x}^{2}\hat{y}^{14}+\\
&-q^{17}(1+q^{2})\hat{x}^{3}\hat{y}^{16}+q^{21}\hat{x}^{3}\hat{y}^{18}-q^{21}(1+q^{2}+q^{4}+q^{6})\hat{x}^{3}\hat{y}^{20}-q^{44}\hat{x}^{4}\hat{y}^{26} .  \label{A-torusknots}
\end{split}
\ee
These results, together with results for $7_1$ and $9_1$ knots, are summarized in the attached supplementary Mathematica file. These dual quantum A-polynomials are quantum versions of, and in the limit $q\to 1$ reduce (possibly up to some simple factor) to, classical dual extremal A-polynomials introduced in \cite{Garoufalidis:2015ewa}. Note that maximal colored polynomials and dual quantum A-polynomial for the trefoil correspond to $m=4$ in (\ref{Punnorm-m}) and (\ref{calAxyP-1}), however all other A-polynomials for torus knots are more complicated than those discussed in section \ref{sub:Knots-with-single}. Moreover, dual minimal A-polynomials for torus knots are of the nonhomogeneous form (coefficients in (\ref{eq:nonhom-Y}) include $\mathcal{A}_{l,0} \neq 0$ for some $l$) discussed in general in section \ref{sub:Generalisation-to-knots}. In what follows we present in detail a combinatorial model associated to the minimal A-polynomial for the trefoil knot and determine corresponding LMOV invariants. Moreover, both for trefoil and for $5_1, 7_1$ and $9_1$ knots, in appendix \ref{app:lmov} 
we illustrate that $Q_{r}(q)$ are indeed divisible by $[r]_{q^2}$ -- which is a consequence of the structure of associated combinatorial models -- and so quantum LMOV invariants, identified as coefficients of $N_{r}(q)$, are indeed integer.

Let us construct a combinatorial model for minimal invariants for the trefoil knot $3_1$, following section \ref{sub:Generalisation-to-knots}. From the form of $\widehat{\mathcal{A}}^{-}_{3_1} (\widehat{x},\widehat{y},q)$ in (\ref{A-torusknots}) it follows that (\ref{eq:nonhom-Y}) takes form
\begin{equation}
\begin{split}
\mathcal{A}(x,Y(x,q),q)=&1-Y(x,q)-q^{-1}x+qxY(x,q)+\\ 
&-qxY(x,q)^{(2;q^{2})}-q^{3}xY(x,q)^{(2;q^{2})}-q^{6}x^{2}Y(x,q)^{(3;q^{2})}=0,   \label{eq:Ainhom31}
\end{split}
\end{equation}
so that $\mathcal{A}_{1,0}=1$. We focus first on the homogeneous version of this equation that does not include the term $q^{-1}x$, and construct $T^{hom}=\bigcup_n T_{n}^{hom}$ as described in section \ref{sub:Combinatorial-construction for coefficients of Y}. Since $I=\sum_{l\geq1,m\geq1,j}\left|\mathcal{A}_{l,m,j}\right|=4$, we consider an alphabet of four letters and corresponding four one-letter words, which we assign to terms in the homogeneous version of (\ref{eq:Ainhom31}), see table \ref{table:letters}.

\begin{table}[h]
\begin{centering}
\caption{Correspondence between one-letter words and terms in $\mathcal{A}^{hom}(x,Y(x,q),q)$.}
\label{table:letters}
\par\end{centering}

\centering{}%
\begin{tabular}{|c|c|}
\hline 
Term in $\mathcal{A}^{hom}(x,Y(x,q),q)$ & One-letter word\tabularnewline
\hline 
\hline 
$qxY(x,q)$ & $\mu$\tabularnewline
\hline 
$-qxY(x,q)^{(2;q^{2})}$ & $\nu$\tabularnewline
\hline 
$-q^{3}xY(x,q)^{(2;q^{2})}$ & $\xi$\tabularnewline
\hline 
$-q^{6}x^{2}Y(x,q)^{(3;q^{2})}$ & $o$\tabularnewline
\hline 
\end{tabular}
\end{table}

Now, starting from $T_{0}^{hom}=\{[\,]\}$, we construct recursively sets $T_n$, for example
\begin{align}
T_{1}^{hom}=&\{[\mu],-[\nu],-[\xi^{*3}]\} = \left\{\young(\mu),-\young(\nu),-\young(\xi,\xi,\xi)\right\}\\
T_{2}^{hom}=&\{[\mu,\mu],-[\mu,\nu],-[\mu,\xi^{*3}],-[\nu,\mu],[\nu,\nu],[\nu,\xi^{*3}],-[\nu,\nu^{*2}\mu],[\nu,\nu^{*3}],[\nu,\nu^{*2}\xi^{*3}],\nonumber \\  
            &-[\xi^{*3},\mu],[\xi^{*3},\nu],[\xi^{*3},\xi^{*3}],-[\xi^{*3},\xi^{*2}\mu],[\xi^{*3},\xi^{*2}\nu],[\xi^{*3},\xi^{*5}],-[\varepsilon,o^{*6}]\} = \\
       = &\left\{ \begin{array}{l} 
                     \young(\mu\mu),-\young(\mu\nu),-\young(:\xi,:\xi,\mu\xi),-\young(\nu\mu),\young(\nu\nu),\young(:\xi,:\xi,\nu\xi),-\young(:\mu,:\nu,\nu\nu),\young(:\nu,:\nu,\nu\nu),\young(:\xi,:\xi,:\xi,:\nu,\nu\nu) \\
                     -\young(\xi:,\xi:,\xi\mu),\young(\xi:,\xi:,\xi\nu),\young(\xi\xi,\xi\xi,\xi\xi),-\young(\xi\mu,\xi\xi,\xi\xi),\young(\xi\nu,\xi\xi,\xi\xi),\young(:\xi,:\xi,\xi\xi,\xi\xi,\xi\xi),-\young(:o,:o,:o,:o,:o,\varepsilon o)
                     \end{array}\right\} \nonumber
\end{align}

Subsequently we construct sets $T^{nonh}=\bigcup_n T^{nonh}_n$. We fix $T_{0}^{nonh}=T_{0}^{hom}=\{[\,]\}$, and since $J=\sum_{l\in\Lambda,j}\left|\mathcal{A}_{l,0,j}\right|=1$,
we introduce an additional one-letter word $\alpha$ corresponding to the nonhomogeneous term $-q^{-1}x$ in (\ref{eq:Ainhom31}), which is lexicographically smaller than $\mu,\nu,\xi,o$, as demanded in section \ref{sub:Generalisation-to-knots}. We then define a new sentence according to (\ref{eq:new sentence nonhomogeneous case})
\begin{equation}
s(1,-1,\alpha)=-[\alpha^{*\left(-1\right)}]=-[\bar{\alpha}].
\end{equation} 
Since $n=l=1$ we have $T_{1}^{aux1}=\left\{ -[\bar{\alpha}]\right\}$. To construct $T_{1}^{aux2}$ we need to consider all partitions with $k_{0}=0$ for the one letter word $\mu$, and $k_{0}+k_{1}=0$ for $\nu$ and $\xi$ (note that for $o$ we would have to consider $k_{0}+k_{1}+k_{2}=-1$, as the ``$o$-recursion'' starts from two word sentences). For $s(1,-1,\alpha)$ we have $k_{i}=1$, so $T_{1}^{aux2}=\left\{ \,\right\}$ and therefore
\be
T_{1}^{nonh}=T_{1}^{hom}\cup T_{1}^{aux1}\cup T_{1}^{aux2}= \left\{[\mu],-[\nu],-[\xi^{*3}],-[\alphabar]\right\} = \left\{\young(\mu),-\young(\nu),-\young(\xi,\xi,\xi),-\young(\alphabar)\right\}.     \label{eq:Tnonh1}
\ee

In the second step of the recursion we determine $T_{2}^{nonh}$. Now $n=2\notin\Lambda$, so $T_{2}^{aux1}=\left\{ \,\right\}$. On the other hand the partition $k_{0}=1$ for the one-letter word $\mu$ corresponds to $s_{k_{0}}=s(1,-1,\alpha)=-[\bar{\alpha}]$ so there is a new sentence
\begin{equation}
s(1,1,1;s_{k_{0}}=-[\bar{\alpha}];\mu)=[\varepsilon]^{*\left(1-1\right)}*[\mu^{*1}]*s_{k_{0}}=-[\mu,\bar{\alpha}].
\end{equation}
For $\nu$ we need to consider partitions $k_{0}+k_{1}=1$, so $s(1,-1,\alpha)=-[\bar{\alpha}]$ appears once as $s_{k_{0}}$ (for $k_{0}=1,\ k_{1}=0$) and once as $s_{k_{1}}$ (for $k_{0}=0,\ k_{1}=1$)
\be
\begin{split}
s(1,2,1;s_{k_{0}}=-[\bar{\alpha}],s_{k_{1}}=[\,];\nu)=&-[\varepsilon]^{*\left(1-1\right)}*[\nu^{*1}]*s_{k_{0}}=[\nu,\bar{\alpha}],  \label{eq:nu1forTaux}\\
s(1,2,1;s_{k_{0}}=[\,],s_{k_{1}}=-[\bar{\alpha}];\nu)=&-[\varepsilon]^{*\left(1-1\right)}*[\nu^{*1}]*\left([\nu^{*2}]^{*k_{1}}\vee s_{k_{1}}\right)=[\nu,\nu^{*2}\bar{\alpha}].  
\end{split}
\ee
and similarly for $\xi$ we get
\be
\begin{split}
s(1,2,3;s_{k_{0}}=-[\bar{\alpha}],s_{k_{1}}=[\,];\xi)=&-[\varepsilon]^{*\left(1-1\right)}*[\xi^{*3}]*s_{k_{0}}=[\xi^{*3},\bar{\alpha}],  \label{eq:xi1forTaux}\\
s(1,2,3;s_{k_{0}}=[\,],s_{k_{1}}=-[\bar{\alpha}];\xi)=&-[\varepsilon]^{*\left(1-1\right)}*[\xi^{*3}]*\left([\xi^{*2}]^{*k_{1}}\vee s_{k_{1}}\right)=[\xi^{*3},\xi^{*2}\bar{\alpha}]. 
\end{split}
\ee
For $o$ we would need to consider $k_{0}+k_{1}+k_{2}=0$, so $s(1,-1,\alpha)=-[\bar{\alpha}]$ does not appear, thus
\be
T_{2}^{aux2}=\left\{-[\mu,\bar{\alpha}],[\nu,\bar{\alpha}],[\nu,\nu^{*2}\bar{\alpha}],[\xi^{*3},\bar{\alpha}],[\xi^{*3},\xi^{*2}\bar{\alpha}] \right\} = \left\{-\young(\mu\alphabar),\young(\nu\alphabar),\young(:\alphabar,:\nu,\nu\nu),\young(\xi:,\xi:,\xi\alphabar),\young(\xi:,\xi\alphabar,\xi\xi)\right\}
\ee
and finally we get
\begin{equation}\label{eq:Tnonh2}
\begin{split}
T_{2}^{nonh}=&T_{2}^{hom}\cup T_{2}^{aux1}\cup T_{2}^{aux2}=  \\
=&\{-[\mu,\bar{\alpha}],[\mu,\mu],-[\mu,\nu],-[\mu,\xi^{*3}],[\nu,\bar{\alpha}],-[\nu,\mu],[\nu,\nu],[\nu,\xi^{*3}],[\nu,\nu^{*2}\bar{\alpha}],\\
&-[\nu,\nu^{*2}\mu],[\nu,\nu^{*3}],[\nu,\nu^{*2}\xi^{*3}],[\xi^{*3},\bar{\alpha}],-[\xi^{*3},\mu],[\xi^{*3},\nu],[\xi^{*3},\xi^{*3}],\\
&[\xi^{*3},\xi^{*2}\bar{\alpha}],-[\xi^{*3},\xi^{*2}\mu],[\xi^{*3},\xi^{*2}\nu],[\xi^{*3},\xi^{*5}],-[\varepsilon,o^{*6}]\}=  \\
=&\left\{ \begin{array}{l}            -\young(\mu\alphabar),\young(\mu\mu),-\young(\mu\nu),-\young(:\xi,:\xi,\mu\xi),\young(\nu\alphabar),-\young(\nu\mu),\young(\nu\nu),\young(:\xi,:\xi,\nu\xi),\\
\\
\young(:\alphabar,:\nu,\nu\nu),-\young(:\mu,:\nu,\nu\nu),\young(:\nu,:\nu,\nu\nu),\young(:\xi,:\xi,:\xi,:\nu,\nu\nu),\young(\xi:,\xi:,\xi\alphabar),-\young(\xi:,\xi:,\xi\mu),\young(\xi:,\xi:,\xi\nu),\young(\xi\xi,\xi\xi,\xi\xi),\\
\\
\young(\xi\alphabar,\xi\xi,\xi\xi),-\young(\xi\mu,\xi\xi,\xi\xi),\young(\xi\nu,\xi\xi,\xi\xi),\young(:\xi,:\xi,\xi\xi,\xi\xi,\xi\xi),-\young(:o,:o,:o,:o,:o,\varepsilon o).
\end{array}\right\}
\end{split}
\end{equation}

Furthermore, according to (\ref{eq:nonhprimary}), we pick primary sentences  $T^{nonh,0}=\bigcup_n T^{nonh,0}_n$ from $T^{nonh}$. In particular for $n=1,2$ we find
\begin{align}\label{eq:Tnonh,0}
T_{1}^{nonh,0}=&\{[\mu],-[\nu],-[\xi^{*3}],-[\alphabar] \}= \left\{\young(\mu),-\young(\nu),-\young(\xi,\xi,\xi),-\young(\alphabar)\right\} \nonumber \\
T_{2}^{nonh,0}=&\{[\nu,\nu^{*2}\bar{\alpha}],-[\nu,\nu^{*2}\mu],[\nu,\nu^{*3}],[\nu,\nu^{*2}\xi^{*3}],[\xi^{*3},\xi^{*2}\bar{\alpha}],\nonumber\\
&-[\xi^{*3},\xi^{*2}\mu],[\xi^{*3},\xi^{*2}\nu],[\xi^{*3},\xi^{*5}],-[\varepsilon,o^{*6}]\}=\\
=&\left\{\young(:\alphabar,:\nu,\nu\nu),-\young(:\mu,:\nu,\nu\nu),\young(:\nu,:\nu,\nu\nu),\young(:\xi,:\xi,:\xi,:\nu,\nu\nu),\young(\xi\alphabar,\xi\xi,\xi\xi),-\young(\xi\mu,\xi\xi,\xi\xi),\young(\xi\nu,\xi\xi,\xi\xi),\young(:\xi,:\xi,\xi\xi,\xi\xi,\xi\xi),-\young(:o,:o,:o,:o,:o,\varepsilon o) \right\}    \nonumber
\end{align}
Now we treat $T^{nonh,0}$ as an alphabet and construct a language $\left(T^{nonh,0}\right)^{*}$. However recall that this time $T^{nonh}\ncong\left(T^{nonh,0}\right)^{*}$. Let us show this explicitly. We define $\left(T^{nonh,0}\right)_{n}^{*}$ as a subset of $\left(T^{nonh,0}\right)^{*}$
whose elements, when mapped by $\varphi$ defined in (\ref{eq:phidef})
and (\ref{eq:phiconc}), consist of $n$ words. For $n=1,2$ we have
\begin{align}
\label{eq:Tnonh,0*} 
\left(T^{nonh,0}\right)^{*}_{1}\overset{\varphi}{\longrightarrow}&T_{1}^{nonh},\\ \left(T^{nonh,0}\right)^{*}_{2}\overset{\varphi}{\longrightarrow}&T_{2}^{nonh}\cup\{[\bar{\alpha},\bar{\alpha}],-[\bar{\alpha},\mu],[\bar{\alpha},\nu],[\bar{\alpha},\xi^{*3}]\}=\nonumber\\ =&T_{2}^{nonh}\cup\left\{\young(\alphabar\alphabar),-\young(\alphabar\mu),\young(\alphabar\nu),\young(:\xi,:\xi,\alphabar\xi) \right\}\neq T_{2}^{nonh}.\nonumber 
\end{align}
Following section \ref{sub:Generalisation-to-knots} we define a set of Lyndon words $T^{nonh,L}$ in the language $\left(T^{nonh,0}\right)^{*}$ and $T_{n}^{nonh,L}=\left(T^{nonh,0}\right)_{n}^{*}\cap T^{nonh,L}$. For $n=1,2$ we obtain
\begin{align}\label{eq:Tnonh,L}
T_{1}^{nonh,L}=&T_{1}^{nonh,0}\\
T_{2}^{nonh,L}=&T_{2}^{nonh,0}\cup\{-[\bar{\alpha}]*[\mu],[\bar{\alpha}]*[\nu],[\bar{\alpha}]*[\xi^{*3}],-[\mu]*[\nu],-[\mu]*[\xi^{*3}],[\nu]*[\xi^{*3}]\}=\nonumber\\
=&T_{2}^{nonh,0}\cup\left\{-\young(\alphabar)*\young(\mu),\young(\alphabar)*\young(\nu),\young(\alphabar)*\young(\xi,\xi,\xi),-\young(\mu)*\young(\nu),-\young(\mu)*\young(\xi,\xi,\xi),\young(\nu)*\young(\xi,\xi,\xi) \right\}
\end{align}
In the next step we define $T^{nonh,L,+}$ according to (\ref{eq:T^L,+-1})
and $T_{n}^{nonh,L,+}=\left(T^{nonh,0}\right)_{n}^{*}\cap T^{nonh,L,+}$
\be
\begin{split}
T_{1}^{nonh,L,+}=&T_{1}^{nonh,L}\\
T_{2}^{nonh,L,+}=&T_{2}^{nonh,L}\cup\{[\bar{\alpha}]*[\bar{\alpha}],[\nu]*[\nu],[\xi^{*3}]*[\xi^{*3}]\}=  \\
=&T_{2}^{nonh,L}\cup\left\{\young(\alphabar)*\young(\alphabar),\young(\nu)*\young(\nu),\young(\xi,\xi,\xi)*\young(\xi,\xi,\xi) \right\}   \label{eq:T^Lnonh}
\end{split}
\ee
In order to adjust $\left(T^{nonh,0}\right)^{*}$ to $T^{nonh}$ we define a relation $\sim$ as explained in section \ref{sub:Generalisation-to-knots}. This relation trivialises some elements of $\left(T^{nonh,0}\right)_{1}^{*}$ and $\left(T^{nonh,0}\right)_{2}^{*}$ 
\be 
[\,]\sim-[\bar{\alpha}]*[\mu] \sim [\bar{\alpha}]*[\nu]\sim[\bar{\alpha}]*[\xi^{*3}], \qquad 
-[\bar{\alpha}] \sim [\bar{\alpha}]*[\bar{\alpha}] .       \label{eq:trivialisation}
\ee
Now we define $\tilde{\varphi}$ that maps the conjugacy class in $\nicefrac{\left(T^{nonh,0}\right)^{*}}{\sim}$ to the image of its shortest representative under $\varphi$
\be
\{[\,],-[\bar{\alpha}]*[\mu],[\bar{\alpha}]*[\nu],[\bar{\alpha}]*[\xi^{*3}]\} \overset{\tilde{\varphi}}{\longmapsto}[\,], \qquad   
\{-[\bar{\alpha}],[\bar{\alpha}]*[\bar{\alpha}]\}\overset{\tilde{\varphi}}{\longmapsto}-[\bar{\alpha}].    \label{eq:phitilde1}
\ee
For classes with one element $\tilde{\varphi}$ reduces to the action of $\varphi$ on the representative, for example
\begin{equation}
\{-[\mu]*[\nu]\}\overset{\tilde{\varphi}}{\longmapsto}-[\mu,\nu]=\varphi\left(-[\mu]*[\nu]\right).
\end{equation}
From (\ref{eq:phitilde1}) we have $\nicefrac{\left(T^{nonh,0}\right)^{*}}{\sim}\overset{\tilde{\varphi}}{\longrightarrow}T^{nonh}$ and in particular 
\be 
\nicefrac{\left(T^{nonh,0}\right)^{*}_{1}}{\sim}\overset{\tilde{\varphi}}{\longrightarrow} T_{1}^{nonh}, \qquad \nicefrac{\left(T^{nonh,0}\right)^{*}_{2}}{\sim}\overset{\tilde{\varphi}}{\longrightarrow} T_{2}^{nonh}.   \label{eq:final isomorphism}
\ee
Now we define $T^{L}$ according to (\ref{eq:T^L nonhomogeneous case}) and $T_{n}^{L} = \tilde{\varphi}\left(\nicefrac{T_{n}^{nonh,L}}{\sim}\right)$, in particular
\be
\begin{split} 
T_{1}^{L}=&T_{1}^{nonh,0}\\
T_{2}^{L}=&T_{2}^{nonh,0}\cup\{-[\mu,\nu],-[\mu,\xi^{*3}],[\nu,\xi^{*3}]\}=T_{2}^{nonh,0}\cup\left\{-\young(\mu\nu),-\young(:\xi,:\xi,\mu\xi),\young(:\xi,:\xi,\nu\xi)     \right\}   \label{eq:T^L final} 
\end{split}
\ee
and then determine $T_{r}^{L,+}$ and $T^{L,+}$, in particular 
\be  
T_{1}^{L,+}=T_{1}^{L},  \qquad T_{2}^{L,+}=T_{2}^{L}\cup\{[\nu,\nu],[\xi^{*3},\xi^{*3}]\}=T_{2}^{L}\cup\left\{\young(\nu\nu),\young(\xi\xi,\xi\xi,\xi\xi)\right\}.   \label{eq:T^L+ final}
\ee
Counting boxes in these sets and including signs, we determine $Q_{r}(q)$ and $N_{r}(q)$ as in (\ref{eq:Qr final})
\be
\begin{split}
Q_{1}(q) &= \textrm{sgn}\left(-[\bar{\alpha}]\right)q^{\textrm{wt}(-[\bar{\alpha}])}+\textrm{sgn}\left([\mu]\right)q^{\textrm{wt}([\mu])}+\textrm{sgn}\left([-\nu]\right)q^{\textrm{wt}(-[\nu])}+\textrm{sgn}\left(-[\xi^{*3}]\right)q^{\textrm{wt}(-[\xi^{*3}])}  \\ 
   &=-q^{-1}+q-q-q^{3}=-q^{-1}-q^{3}=N_{1}(q),\nonumber\\ 
Q_{2}(q) &= q^2+q^4+q^6+q^8,\qquad N_{2}(q)=\frac{Q_{2}(q)}{[2]_{q^{2}}}= q^2+q^6.\label{eq:N2 final} 
\end{split}
\ee
We conclude that quantum LMOV invariants for $r=1,2$ take form
\be
N_{1,-2}=N_{1,2}=-1,\qquad N_{2,1}=N_{2,5}=1,
\ee
Results for $Q_{r}(q)$ and $N_{r}(q)$ up to $r=9$, which in particular confirm integrality of quantum LMOV invariants, are given in table \ref{tab:31 numerical results}.



\acknowledgments{We thank Markus Reineke and Marko Sto$\check{\text{s}}$i$\acute{\text{c}}$ for discussions on these and related topics. This work is supported by the ERC Starting Grant no. 335739 \emph{``Quantum fields and knot homologies''} funded by the European Research Council under the European Union's Seventh Framework Programme, and the Foundation for Polish Science.}


\newpage

\appendix

\section{Relation between $Y_{n}(q)$ and $Q_{r}(q)$\label{sec:ConnectionY-QandRecursionN}}

In this appendix we prove the relation (\ref{eq:Ydependence_onQ}) by a direct computation. From (\ref{eq:Y def}) 
\begin{equation}
n!\, Y_{n}(q)=\partial_{x}^{n}Y(x,q)\Big|_{x=0} =\partial_{x}^{n}\prod_{r=1}^{\infty} \prod_{l=0}^{r-1}
\prod_{j=j_{min}}^{j_{max}}\Big(1-x^{r}q^{j+1+2l}\Big)^{-N_{r,j}}\Big|_{x=0}.\label{eq:nth deriv of Y(x,q)}
\end{equation}
Let us write $Y(x,q)=\prod_{r=1}^{\infty} \tilde{Y}_r(x,q)$ and introduce 
\be
\tilde{Y}_{r}(x,q) = \prod_{l=1}^{r-1} \tilde{Y}_{r,l}(x,q),\quad
\tilde{Y}_{r,l}(x,q) = \prod_{j=j_{min}}^{j_{max}} \tilde{Y}_{r,l,j}(x,q), \quad 
\tilde{Y}_{r,l,j}(x,q) = \big(1-x^{r}q^{j+1+2l}\big)^{-N_{r,j}}.
\ee
First note that
\be
\partial_{x}\tilde{Y}_{r,j,l}(x,q) = \left(-N_{r,j}\right)\left(1-x^{r}q^{j+1+2l}\right)^{-N_{r,j}-1}\left(-rx^{r-1}q^{j+1+2l}\right) \label{eq:derivative of Yrjl}
\ee
is non-zero at $x=0$ only for $r=1$. To get a non-zero result for $r>1$ we need to take additional $r-1$ derivatives
of $\left(-rx^{r-1}q^{j+1+2l}\right)$
\begin{equation}
\partial_{x}^{r}\tilde{Y}_{r,l,j}(x,q)\Big|_{x=0}=\partial_{x}^{r}\left(1-x^{r}q^{j+1+2l}\right)^{-N_{r,j}}\Big|_{x=0}=r!N_{r,j}q^{j+1+2l}.
\end{equation}
Furthermore, only derivatives of multiple orders in $r$ give non-zero contribution at $x=0$
\begin{equation}
\begin{split}\label{eq:rv derivative of Yrjl}
&\partial_{x}^{rv}\tilde{Y}_{r,l,j}(x,q)\Big|_{x=0}=\frac{1}{v!}\left(\begin{array}{c}rv\\r,r,\ldots,r\end{array}\right)\left[r!N_{r,j}q^{j+1+2l}\right]\left[r!\left(N_{r,j}+1\right)q^{j+1+2l}\right]\times\ldots\\
&\qquad \ldots\times\left[r!\left(N_{r,j}+v-2\right)q^{j+1+2l}\right]\left[r!\left(N_{r,j}+v-1\right)q^{j+1+2l}\right]= \frac{\left(rv\right)!}{v!}N_{r,j}^{(v)}\left(q^{j+1+2l}\right)^{v}
\end{split}
\end{equation}
where $N_{r,j}^{(v)}=N_{r,j}\left(N_{r,j}+1\right)\ldots\left(N_{r,j}+v-1\right)$ is a Pochhammer symbol, and 
\be
\frac{1}{v!}\left(\begin{array}{c}
rv\\
r,r,\ldots,r
\end{array}\right)=\frac{\left(rv\right)!}{v!(r!)^{v}}
\ee 
is the number of ways of dividing $rv$ elements into $v$ indistinguishable groups of $r$ elements each. Generalizing the Pochhammer symbol to polynomials as in (\ref{eq:Pochhammerforpoly}), we can write 
\begin{equation} 
\begin{split}\label{eq:rv derivative of Yrl}
&\partial_{x}^{rv}\tilde{Y}_{r,l}(x,q)\Big|_{x=0}=\underset{ru_{j_{min}}+\ldots+ru_{j_{max}}=rv}{\sum}\left(\begin{array}{c}rv\\ru_{j_{min}},\ldots,ru_{j_{max}}\end{array}\right)\underset{j=j_{min}}{\stackrel{j_{max}}{\prod}}\frac{\left(ru_{j}\right)!}{u_{j}!}N_{r,j}^{(u_{j})}q^{(j+1+2l)u_{j}}=\\ 
&\  =\frac{\left(rv\right)!}{v!}\underset{ru_{j_{min}}+\ldots+ru_{j_{max}}=rv}{\sum}\left(\begin{array}{c}v\\u_{j_{min}},\ldots,u_{j_{max}}\end{array}\right)\underset{j=j_{min}}{\stackrel{j_{max}}{\prod}}N_{r,j}^{(u_{j})}q^{(j+1+2l)u_{j}} = \frac{\left(rv\right)!}{v!}q^{2lv}N_{r}(q)^{(v)} 
\end{split} 
\end{equation}
and analogously
\begin{equation} 
\begin{split}\label{eq:rv derivative of Yr}
&\partial_{x}^{rv}\tilde{Y}_{r}(x,q)\Big|_{x=0} = \underset{rt_{0}+rt_{1}+\ldots+rt_{r-1}=rv}{\sum}\left(\begin{array}{c}
rv\\
rt_{0},rt_{1},\ldots,rt_{r-1}
\end{array}\right)\underset{l=1}{\stackrel{r-1}{\prod}}\frac{\left(rt_{l}\right)!}{t_{l}!}\left(q^{2l}\right)^{t_{l}}N_{r}(q)^{(t_{l})}=\\ 
&\  =\frac{\left(rv\right)!}{v!}\underset{t_{0}+t_{1}+\ldots+t_{r-1}=v}{\sum}\left(\begin{array}{c}
v\\
t_{0},t_{1},\ldots,t_{r-1}
\end{array}\right)\underset{l=1}{\stackrel{r-1}{\prod}}\left(N_{r}(q)q^{2l}\right)^{(t_{l})}=\frac{\left(rv\right)!}{v!}\left(N_{r}(q)[r]_{q^{2}}\right)^{(v)}, 
\end{split} 
\end{equation}
where $[r]_{q^{2}}=\frac{1-q^{2r}}{1-q^{2}}=\sum_{l=0}^{r-1} q^{2l}.$
Note that the expansion of $\big(\sum_{l=1}^{r-1} N_{r}(q)q^{2l}\big)^{(v)}$ in terms of $(N_{r}(q)q^{2l})^{(t_{l})}$ is the same as in an ordinary multinomial formula.

Once we found $\partial_{x}^{rv}\tilde{Y}_{r}(x,q)\big|_{x=0}$ we can calculate $\partial_{x}^{n}Y(x,q)\big|_{x=0}$ using the Leibniz rule. Note that the $n$-th derivative $\partial_{x}^{n}$ in $\partial_{x}^{n}  \prod_r \tilde{Y}_{r}(x,q)\big|_{x=0}$, when acting on $\tilde{Y}_{r}(x,q)$ with $r>n$, always gives $0$. 
Therefore 
\begin{equation} 
\begin{split}    
Y_n(q)  = & \frac{1}{n!} \partial_{x}^{n}Y(x,q)\Big|_{x=0} = \frac{1}{n!} \partial_{x}^{n}\prod_{r=1}^n \left.\tilde{Y}_{r}(x,q)\right|_{x=0}=\\ =& \frac{1}{n!} \sum_{1v_{1}+2v_{2}+\ldots+nv_{n}=n}\left(\begin{array}{c}
n\\
1v_{1},2v_{2},\ldots,nv_{n}
\end{array}\right)\prod_{r=1}^n \frac{\left(rv_{r}\right)!}{v_{r}!}\left(N_{r}(q)[r]_{q^{2}}\right)^{(s_{r})}=  \label{eq:n derivative of Y} \\ 
=&
\sum_{1v_{1}+2v_{2}+\ldots+nv_{n}=n}\left(\prod_{r=1}^n \frac{Q_{r}(q)^{(v_{r})}}{v_{r}!}\right), 
\end{split} 
\end{equation}
which proves the relation (\ref{eq:Ydependence_onQ}).


\section{Relation to  the model in \cite{Rei12}  \label{sec:Duality us-Reineke}}

Extremal colored polynomials for twist knots labeled by non-negative $m$, discussed in section \ref{sub:Knots-with-single}, are closely related to generating series associated to $m$-loop quivers analyzed in \cite{Rei12}. For this reason the combinatorial model presented in section \ref{sub:Knots-with-single} is equivalent to the model describing combinatorics of the degenerate Cohomological Hall algebra of the $m$-loop quiver, introduced in \cite{Rei12}. In this section we present precise relation between these two models. 

Let us denote by $T^R_n$ the set referred to as $T_n$ in \cite{Rei12}, which consists of partitions $\lambda=\left(\lambda_{1},\ldots,\lambda_{n}\right)$
such that $0\leq\lambda_{1}\leq\ldots\leq\lambda_{n}$ and $\lambda_{i}\leq(m-1)(i-1)$ for all $i=1,\ldots,n$. There is a bijection  $\Phi: T^R_n \to T_n$ between $T^R_n$ and the set $T_n$ in our model, defined by 
\begin{equation}
\Phi(\lambda)=(-1)^{(m+1)n}\left[\phi_{1},\phi_{2},\ldots,\phi_{n}\right],  \qquad   \phi_{i}=2\left((m-1)(i-1)-\lambda_{i}\right)+1.
\end{equation}
Note that $\phi_{i+1}-\phi_{i}\leq2(m-1)$, and the inequality $\lambda_{i}\leq(m-1)(i-1)$ ensures that all elements of $\Phi(\lambda)$ are positive and $\phi_{1}=1$, so that indeed $\Phi(\lambda)\in T_{n}$. $\Phi$ is a surjection because every $n$-element list $\phi=(-1)^{(m+1)n}\left[\phi_{1},\phi_{2},\ldots,\phi_{n}\right]$,
such that $\phi_{1}=1$ and $\phi_{i}\geq1$ for $i=2,\ldots,n$ and $\phi_{i+1}-\phi_{i}\leq2(m-1)$, can be written as $\phi_{i}=2\left((m-1)(i-1)-\lambda_{i}\right)+1$,
where $0\leq\lambda_{1}\leq\ldots\leq\lambda_{n}$ and $\lambda_{i}\leq(m-1)(i-1)$. $\Phi$ is also clearly an injection, so it is a bijection between $T^{\textrm{R}}$ and $T$ that preserves the length $n$. Moreover the sets $T^{0}$, $T^{L}$ and $T^{L,+}$ are defined in section \ref{sub:Knots-with-single} in the same way as analogous sets in \cite{Rei12}, which we refer to as $T^{R,0}$, $T^{R,L}$ and $T^{R,L,+}$, therefore $\Phi$ is directly generalized to a bijection between $T^{R,0}$, $T^{R,L}$, $T^{R,L,+}$ and $T^{0}$, $T^{L}$, $T^{L,+}$ respectively. 

Furthermore, a weight of a partition $\lambda\in T^R$ is defined in \cite{Rei12} by 
$\textrm{wt}^{R}(\lambda)=(m-1) {n \choose 2} -\left|\lambda\right|$, 
where $\left|\lambda\right|=\lambda_{1}+\cdots+\lambda_{n}$, therefore
\begin{equation}
\begin{split}
\textrm{wt}(\phi(\lambda))=\overset{n}{\underset{i=1}{\sum}}\phi_{i} = \overset{n}{\underset{i=1}{\sum}}2\left((m-1)(i-1)-\lambda_{i}\right)+1 
= 2\textrm{wt}^{R}(\lambda)+n.
\end{split}
\end{equation}
Since $Q_{r}^{R}(q)=\sum_{\lambda\in T_{r}^{R,L,+}} q^{\textrm{wt}^{R}(\lambda)}$, it follows that
\begin{equation} 
\begin{split} 
Q_{r}(q)&=\sum_{\phi\in T_{r}^{L,+}} \textrm{sgn}\left(\phi\right)q^{\textrm{wt}(\phi)} = \sum_{\lambda\in T_{r}^{R,L,+}} (-1)^{r(m+1)}q^{2\textrm{wt}^{R}(\lambda)+r} = (-1)^{r(m+1)}q^{r}Q_{r}^{R}(q^{2}).
\end{split} 
\end{equation}
It is proven in \cite{Rei12} that $Q_{r}^{R}(q)$ is divisible by $\left[r\right]_{q}=\frac{1-q^{r}}{1-q}$ (because $\zeta=1$ is the sole $r$-th root of unity for which
$Q_{r}^{\textrm{R}}(q)\neq0$), and this quotient is proportional to quantized Donaldson-Thomas invariants for the $m$-loop quiver
\begin{equation}
\textrm{DT}_{r}^{(m)}(q)=q^{1-r}\frac{1}{\left[r\right]_{q}}Q_{r}^{R}(q).\label{eq:Reineke's result}
\end{equation}
Recalling (\ref{eq:Nr(q)}) it follows that
\begin{equation}
N_{r}^{m}(q)=\frac{Q_{r}(q)}{[r]_{q^{2}}}
=(-1)^{r(m+1)}q^{3r-2}\textrm{DT}_{r}^{(m)}(q^{2}).
\end{equation}
Therefore also $N_{r}^{m}(q)=\underset{j}{\sum}N_{r,j}^{m}q^{j+1}$ are polynomials with integer coefficients, which proves the integrality of LMOV invariats discussed in section \ref{sub:Knots-with-single}, labeled by non-negative $m$.




\section{Quantum LMOV invariants for torus knots}  \label{app:lmov}

\begin{table}[H]
\begin{centering}
\caption{Numerical results for $Q_{r}(q)$ and $N_{r}(q)$ for $5_{1}^{+}$ case.}
\label{tab:51max numerical results}
\par\end{centering}

\centering{}%
\begin{tabular}{|c|c|c|}
\hline 
$r$ & $Q_{r}(q)$ & $N_{r}(q)=\nicefrac{Q_{r}(q)}{[r]_{q^{2}}}$ \tabularnewline
\hline 
\hline 
$1$ & $-q^{-1}-q^{3}$ & $-q^{-1}-q^{3}$\tabularnewline
\hline 
$2$ & $\begin{array}{c}
1+q^{-2}+2q^{2}+3q^{4}+3q^{6}\\
+3q^{8}+3q^{10}+2q^{12}+q^{14}+q^{16}
\end{array}$ & $\begin{array}{c}
q^{-2}+2q^{2}+q^{4}+2q^{6}\\
+q^{8}+2q^{10}+q^{14}
\end{array}$\tabularnewline
\hline 
$3$ & $\begin{array}{c}
-q^{-1}-3q-6q^{3}-10q^{5}-15q^{7}-19q^{9}\\
-24q^{11}-26q^{13}-28q^{15}-26q^{17}-25q^{19}\\
-21q^{21}-19q^{23}-14q^{25}-12q^{27}-8q^{29}\\
-7q^{31}-4q^{33}-3q^{35}-q^{37}-q^{39}
\end{array}$ & $\begin{array}{c}
-q^{-1}-2q-3q^{3}-5q^{5}-7q^{7}\\
-7q^{9}-10q^{11}-9q^{13}-9q^{15}\\
-8q^{17}-8q^{19}-5q^{21}-6q^{23}\\
-3q^{25}-3q^{27}-2q^{29}-2q^{31}-q^{35}
\end{array}$\tabularnewline
\hline 
\end{tabular}
\end{table}

\begin{table}[H]
\begin{centering}
\caption{Numerical results for $Q_{r}(q)$ and $N_{r}(q)$ for $7_{1}^{+}$ case.}
\label{tab:71max numerical results}
\par\end{centering}

\centering{}%
\begin{tabular}{|c|c|c|}
\hline 
$r$ & $Q_{r}(q)$ & $N_{r}(q)=\nicefrac{Q_{r}(q)}{[r]_{q^{2}}}$ \tabularnewline
\hline 
\hline 
$1$ & $-q^{-3}-q-q^{5}$ & $-q^{-3}-q-q^{5}$\tabularnewline
\hline 
$2$ & $\begin{array}{c}
q^{-6}+q^{-4}+2q^{-2}+3+4q^{2}+5q^{4}\\
+6q^{6}+6q^{8}+6q^{10}+6q^{12}+5q^{14}\\
+4q^{16}+3q^{18}+2q^{20}+q^{22}+q^{24}
\end{array}$ & $\begin{array}{c}
q^{-6}+2q^{-2}+1+3q^{2}+2q^{4}\\
+4q^{6}+2q^{8}+4q^{10}+2q^{12}\\
+3q^{14}+q^{16}+2q^{18}+q^{22}
\end{array}$\tabularnewline
\hline 
$3$ & $\begin{array}{c}
-q^{-7}-3q^{-5}-6q^{-3}-11q^{-1}-18q\\
-27q^{3}-39q^{5}-51q^{7}-65q^{9}-77q^{11}\\
-90q^{13}-98q^{15}-106q^{17}-107q^{19}\\
-108q^{21}-102q^{23}-98q^{25}-87q^{27}\\
-80q^{29}-67q^{31}-60q^{33}-48q^{35}-42q^{37}\\
-32q^{39}-27q^{41}-19q^{43}-16q^{45}-10q^{47}\\
-8q^{49}-4q^{51}-3q^{53}-q^{55}-q^{57}
\end{array}$ & $\begin{array}{c}
-q^{-7}-2q^{-5}-3q^{-3}-6q^{-1}-9q\\
-12q^{3}-18q^{5}-21q^{7}-26q^{9}\\
30q^{11}-34q^{13}-34q^{15}-38q^{17}\\
-35q^{19}-35q^{21}-32q^{23}-31q^{25}\\
-24q^{27}-25q^{29}-18q^{31}-17q^{33}\\
-13q^{35}-12q^{37}-7q^{39}-8q^{41}\\
-4q^{43}-4q^{45}-2q^{47}-2q^{49}-q^{53}
\end{array}$\tabularnewline
\hline 
\end{tabular}
\end{table}

\begin{table}[H]
\begin{centering}
\caption{Numerical results for $Q_{r}(q)$ and $N_{r}(q)$ for $9_{1}^{+}$ case.}
\label{tab:91max numerical results}
\par\end{centering}

\centering{}%
\begin{tabular}{|c|c|c|}
\hline 
$r$ & $Q_{r}(q)$ & $N_{r}(q)=\nicefrac{Q_{r}(q)}{[r]_{q^{2}}}$ \tabularnewline
\hline 
\hline 
$1$ & $-q^{-5}-q^{-1}-q^{3}-q^{7}$ & $-q^{-5}-q^{-1}-q^{3}-q^{7}$\tabularnewline
\hline 
$2$ & $\begin{array}{c}
q^{-10}+q^{-8}+2q^{-6}+3q^{-4}+4q^{-2}\\
+7q^{2}+8q^{4}+9q^{6}+10q^{8}\\
+10q^{10}+10q^{12}+10q^{14}+9q^{16}\\
+8q^{18}+7q^{20}+5q^{22}+4q^{24}+\\
3q^{26}+2q^{28}+q^{30}+q^{32}
\end{array}$ & $\begin{array}{c}
q^{-10}+2q^{-6}+q^{-4}+3q^{-2}\\
+2+5q^{2}+3q^{4}+6q^{6}\\
+4q^{8}+6q^{10}+4q^{12}+6q^{14}\\
+3q^{16}+5q^{18}+2q^{20}\\
+3q^{22}+q^{24}+2q^{26}+q^{30}
\end{array}$\tabularnewline
\hline 
$3$ & $\begin{array}{c}
-q^{-13}-3q^{-11}-6q^{-9}-11q^{-7}-18q^{-5}\\
-28q^{-3}-42q^{-1}-59q-80q^{3}-103q^{5}\\
-130q^{7}-157q^{9}-187q^{11}-213q^{13}\\
-240q^{15}-260q^{17}-280q^{19}-289q^{21}\\
-298q^{23}-295q^{25}-293q^{27}-279q^{29}\\
-268q^{31}-246q^{33}-230q^{35}-204q^{37}\\
-187q^{39}-161q^{41}-145q^{43}-121q^{45}\\
-107q^{47}-87q^{49}-76q^{51}-59q^{53}\\
-50q^{55}-37q^{57}-31q^{59}-21q^{61}-17q^{63}\\
-10q^{65}-8q^{67}-4q^{69}-3q^{71}-q^{73}-q^{75}
\end{array}$ & $\begin{array}{c}
-q^{-13}-2q^{-11}-3q^{-9}-6q^{-7}-9q^{-5}\\
-13q^{-3}-20q^{-1}-26q-34q^{3}\\
-43q^{5}-53q^{7}-61q^{9}-73q^{11}\\
-79q^{13}-88q^{15}-93q^{17}-99q^{19}\\
-97q^{21}-102q^{23}-96q^{25}-95q^{27}\\
-88q^{29}-85q^{31}-73q^{33}-72q^{35}\\
-59q^{37}-56q^{39}-46q^{41}-43q^{43}\\
-32q^{45}-32q^{47}-23q^{49}-21q^{51}\\
-15q^{53}-14q^{55}-8q^{57}-9q^{59}\\
-4q^{61}-4q^{63}-2q^{65}-2q^{67}-q^{71}
\end{array}$\tabularnewline
\hline 
\end{tabular}
\end{table}

\begin{table}[H]
\begin{centering}
\caption{Numerical results for $Q_{r}(q)$ and $N_{r}(q)$ for $3_{1}^{-}$ case.}
\label{tab:31 numerical results}
\par\end{centering}

\centering{}%
\begin{tabular}{|c|c|c|}
\hline 
$r$ & $Q_{r}(q)$ & $N_{r}(q)=\nicefrac{Q_{r}(q)}{[r]_{q^{2}}}$ \tabularnewline
\hline 
\hline 
$1$ & $-q^{-1}-q^{3}$ & $-q^{-1}-q^{3}$\tabularnewline
\hline 
$2$ & $q^{2}+q^{4}+q^{6}+q^{8}$ & $q^{2}+q^{6}$\tabularnewline
\hline 
$3$ & $-q^{5}-2q^{7}-2q^{9}-2q^{11}-q^{13}-q^{15}$ & $-q^{5}-q^{7}-q^{11}$\tabularnewline
\hline 
$4$ & $\begin{array}{c}
q^{6}+2q^{8}+4q^{10}+5q^{12}+6q^{14}\\
+5q^{16}+4q^{18}+3q^{20}+q^{22}+q^{24}
\end{array}$ & $q^{6}+q^{8}+2q^{10}+q^{12}+2q^{14}+q^{18}$\tabularnewline
\hline 
$5$ & $\begin{array}{c}
-q^{7}-3q^{9}-6q^{11}-10q^{13}-14q^{15}\\
-17q^{17}-18q^{19}-17q^{21}-15q^{23}-11q^{25}\\
-8q^{27}7-5q^{29}-3q^{31}-q^{33}-q^{35}
\end{array}$ & $\begin{array}{c}
-q^{7}-2q^{9}-3q^{11}-4q^{13}-4q^{15}\\
-4q^{17}-3q^{19}-2q^{21}-2q^{23}-q^{27}
\end{array}$\tabularnewline
\hline 
$6$ & $\begin{array}{c}
6q^{8}+4q^{10}+8q^{12}+17q^{14}+26q^{16}\\
+38q^{18}+48q^{20}+57q^{22}+60q^{24}+60q^{26}\\
+55q^{28}+47q^{30}+38q^{32}+28q^{34}+21q^{36}\\
+13q^{38}+9q^{40}+5q^{42}+3q^{44}+q^{46}+q^{48}
\end{array}$ & $\begin{array}{c}
6q^{8}+3q^{10}+4q^{12}+9q^{14}+9q^{16}\\
+12q^{18}+11q^{20}+12q^{22}\\
+7q^{24}+9q^{26}+4q^{28}\\
+4q^{30}+2q^{32}+2q^{34}+q^{38}
\end{array}$\tabularnewline
\hline 
$7$ & $\begin{array}{c}
-q^{9}-4q^{11}-11q^{13}-24q^{15}-44q^{17}\\
-71q^{19}-103q^{21}-137q^{23}-169q^{25}\\
-195q^{27}-211q^{29}-216q^{31}-208q^{33}\\
-192q^{35}-168q^{37}-142q^{39}-114q^{41}\\
-89q^{43}-66q^{45}-49q^{47}-33q^{49}-23q^{51}\\
-14q^{53}-9q^{55}-5q^{57}-3q^{59}-q^{61}-q^{63}
\end{array}$ & $\begin{array}{c}
-q^{9}-3q^{11}-7q^{13}-13q^{15}-20q^{17}\\
-27q^{19}-32q^{21}-35q^{23}\\
-35q^{25}-33q^{27}-29q^{29}-25q^{31}\\
-19q^{33}-16q^{35}-11q^{37}-9q^{39}\\
-5q^{41}-4q^{43}-2q^{45}-2q^{47}-q^{51}
\end{array}$\tabularnewline
\hline 
$8$ & $\begin{array}{c}
q^{11}+4q^{12}+14q^{14}+33q^{16}+70q^{18}\\
+123q^{20}+200q^{22}+290q^{24}+398q^{26}\\
+505q^{28}+613q^{30}+700q^{32}+768q^{34}\\
+801q^{36}+805q^{38}+776q^{40}+723q^{42}\\
+652q^{44}+568q^{46}+484q^{48}+399q^{50}\\
+324q^{52}+253q^{54}+197q^{56}+145q^{58}\\
+108q^{60}+76q^{62}+54q^{64}+35q^{66}+24q^{68}\\
+14q^{70}+9q^{72}+5q^{74}+3q^{76}+q^{78}+q^{80}
\end{array}$ & $\begin{array}{c}
8q^{10}+3q^{12}+10q^{14}+19q^{16}\\
+37q^{18}+53q^{20}+77q^{22}\\
+90q^{24}+109q^{26}+110q^{28}\\
+118q^{30}+106q^{32}+105q^{34}\\
+86q^{36}+81q^{38}+61q^{40}\\
+56q^{42}+39q^{44}+34q^{46}+22q^{48}\\
+20q^{50}+11q^{52}+10q^{54}+5q^{56}\\
+4q^{58}+2q^{60}+2q^{62}+q^{66}
\end{array}$\tabularnewline
\hline 
$9$ & $\begin{array}{c}
-q^{11}-5q^{13}-17q^{15}-47q^{17}-106q^{19}\\
-207q^{21}-363q^{23}-578q^{25}-851q^{27}\\
-1176q^{29}-1536q^{31}-1910q^{33}-2275q^{35}\\
-2602q^{37}-2867q^{39}-3050q^{41}-3141q^{43}\\
-3134q^{45}-3042q^{47}-2872q^{49}-2649q^{51}\\
-2386q^{53}-2108q^{55}-1823q^{57}\\
-1551q^{59}-1292q^{61}-1063q^{63}\\
-855q^{65}-681q^{67}-530q^{69}-408q^{71}\\
-305q^{73}-228q^{75}-163q^{77}-118q^{79}\\
-81q^{81}-56q^{83}-36q^{85}-24q^{87}-14q^{89}\\
-9q^{91}-5q^{93}-3q^{95}-q^{97}-q^{99}
\end{array}$ & $\begin{array}{c}
-q^{11}-4q^{13}-12q^{15}-30q^{17}-59q^{19}\\
-101q^{21}-156q^{23}-215q^{25}-273q^{27}\\
-326q^{29}-364q^{31}-386q^{33}-395q^{35}\\
-386q^{37}-366q^{39}-339q^{41}-306q^{43}\\
-266q^{45}-234q^{47}-194q^{49}\\
-163q^{51}-132q^{53}-108q^{55}\\
-81q^{57}-67q^{59}-47q^{61}-37q^{63}\\
-26q^{65}-20q^{67}-12q^{69}-10q^{71}\\
-5q^{73}-4q^{75}-2q^{77}-2q^{79}-q^{83}
\end{array}$\tabularnewline
\hline 
\end{tabular}
\end{table}

\begin{table}[H]
\begin{centering}
\caption{Numerical results for $Q_{r}(q)$ and $N_{r}(q)$ for $5_{1}^{-}$ case.}
\label{tab:51min numerical results}
\par\end{centering}

\centering{}%
\begin{tabular}{|c|c|c|}
\hline 
$r$ & $Q_{r}(q)$ & $N_{r}(q)=\nicefrac{Q_{r}(q)}{[r]_{q^{2}}}$ \tabularnewline
\hline 
\hline 
$1$ & $-q^{-3}-q-q^{5}$ & $-q^{-3}-q-q^{5}$\tabularnewline
\hline 
$2$ & $\begin{array}{c}
q^{-2}+1+2q^{2}+3q^{4}+3q^{6}\\
+3q^{8}+3q^{10}+2q^{12}+q^{14}+q^{16}
\end{array}$ & $\begin{array}{c}
q^{-2}+2q^{2}+q^{4}+2q^{6}\\
+q^{8}+2q^{10}+q^{14}
\end{array}$\tabularnewline
\hline 
$3$ & $\begin{array}{c}
-q^{-1}-3q-6q^{3}-10q^{5}-14q^{7}\\
-18q^{9}-21q^{11}-23q^{13}-22q^{15}\\
-21q^{17}-17q^{19}-15q^{21}-10q^{23}\\
-8q^{25}-4q^{27}-3q^{29}-q^{31}-q^{33}
\end{array}$ & $\begin{array}{c}
-q^{-1}-2q-3q^{3}-5q^{5}\\
-6q^{7}-7q^{9}-8q^{11}-8q^{13}\\
-6q^{15}-7q^{17}-4q^{19}\\
-4q^{21}-2q^{23}-2q^{25}-q^{29}
\end{array}$\tabularnewline
\hline 
\end{tabular}
\end{table}

\begin{table}[H]
\begin{centering}
\caption{Numerical results for $Q_{r}(q)$ and $N_{r}(q)$ for $7_{1}^{-}$ case.}
\label{tab:71min numerical results}
\par\end{centering}

\centering{}%
\begin{tabular}{|c|c|c|}
\hline 
$r$ & $Q_{r}(q)$ & $N_{r}(q)=\nicefrac{Q_{r}(q)}{[r]_{q^{2}}}$ \tabularnewline
\hline 
\hline 
$1$ & $-q^{-5}-q^{-1}-q^{3}-q^{7}$ & $-q^{-5}-q^{-1}-q^{3}-q^{7}$\tabularnewline
\hline 
$2$ & $\begin{array}{c}
q^{-6}+q^{-4}+2q^{-2}+1+3q^{2}\\
+2q^{4}+4q^{6}+2q^{8}+4q^{10}\\
+2q^{12}+3q^{14}+q^{16}+2q^{18}+q^{22}
\end{array}$ & $\begin{array}{c}
q^{-6}+2q^{-2}+1+3q^{2}+2q^{4}\\
+4q^{6}+2q^{8}+4q^{10}+2q^{12}\\
+3q^{14}+q^{16}+2q^{18}+q^{22}
\end{array}$\tabularnewline
\hline 
$3$ & $\begin{array}{c}
-q^{-7}-3q^{-5}-6q^{-3}-11q^{-1}-18q\\
-27q^{3}-38q^{5}-50q^{7}-62q^{9}\\
-74q^{11}-83q^{13}-91q^{15}-94q^{17}\\
-96q^{19}-91q^{21}-87q^{23}-77q^{25}\\
-70q^{27}-57q^{29}-49q^{31}-37q^{33}\\
-31q^{35}-21q^{37}-17q^{39}-10q^{41}\\
-8q^{43}-4q^{45}-3q^{47}-q^{49}-q^{51}
\end{array}$ & $\begin{array}{c}
-q^{-7}-2q^{-5}-3q^{-3}-6q^{-1}\\
-9q-12q^{3}-17q^{5}-21q^{7}\\
-24q^{9}-29q^{11}-30q^{13}-32q^{15}\\
-32q^{17}-32q^{19}-27q^{21}-28q^{23}\\
-22q^{25}-20q^{27}-15q^{29}-14q^{31}\\
-8q^{33}-9q^{35}-4q^{37}\\
-4q^{39}-2q^{41}-2q^{43}-q^{47}
\end{array}$\tabularnewline
\hline 
\end{tabular}
\end{table}



\newpage

\bibliographystyle{JHEP}
\bibliography{abmodel}

\end{document}